\documentclass[longbibliography,aps,prf,onecolumn,nobibnotes,superscriptaddress,floatfix,tightenlines,showpacs,notitlepage]{revtex4-1}
\usepackage{graphicx}
\usepackage{bm}
\usepackage[normalem]{ulem}
\usepackage{amsfonts}
\usepackage{color}
\usepackage{ulem}
\usepackage{amsmath}    
\usepackage{epsfig}
\usepackage{subfigure}  
\usepackage[breaklinks,colorlinks = true,linkcolor = blue,urlcolor=blue,citecolor=blue]{hyperref}
\usepackage{amssymb}
\usepackage{lineno}
\usepackage{hyperref} 
\usepackage{cancel}
\usepackage{comment}
\usepackage{tikz}

\begin{document}

\title{Is Vorticity Amplification Essential for Anomalous Dissipation \\ and Intermittency in Turbulence?}

\author{Alessandro Chiarini}
\email[]{alessandro.chiarini@polimi.it}
\affiliation{Dipartimento di Scienze e Tecnologie Aerospaziali, Politecnico di Milano, via La Masa 34, 20156 Milano, Italy.}
                          
\date{\today}

\begin{abstract}
The distinct dynamical roles of vorticity amplification (VA) and vortex tilting (VT) in three-dimensional homogeneous isotropic turbulence are investigated using direct numerical simulations of a modified Navier--Stokes system. By selectively suppressing VA while retaining VT, we demonstrate that gradient amplification by VA is strictly required to sustain the classical forward energy cascade. Progressively suppressing VA weakens small-scale velocity gradients, heavily depletes extreme fluctuations, and entirely eliminates the classical energy dissipative anomaly, causing the normalized energy dissipation to decay as $Re_\lambda^{-1}$. In the limit of complete VA suppression, enstrophy emerges as the relevant inviscid invariant. Scale-by-scale budget analyses confirm that the dynamics transition to a purely forward enstrophy cascade, characterized by an anomalous enstrophy dissipation and a robust $E(k) \sim k^{-3}$ intermediate energy spectrum. We verify that these asymptotic scaling limits---including a $Re_\lambda^{-1/2}$ finite-Reynolds-number correction for the enstrophy anomalous dissipation---are universal properties of the VA-suppressed dynamics, independent of whether the large-scale forcing is helical or non-helical. Remarkably, despite the smoothing of the velocity field and the elimination of the energy dissipative anomaly, anomalous structural scaling and a broad multifractal spectrum persist. These results reveal a fundamental mechanistic separation: while VA is responsible for amplifying intense localized fluctuations, geometric reorganization by VT alone is mechanically sufficient to sustain multifractal intermittency.
\end{abstract}
\maketitle

\section{Introduction and aim}

The nonlinear dynamics of the Navier--Stokes equations, and particularly the evolution of the vorticity field, lie at the heart of three-dimensional hydrodynamic turbulence. The intricate couplings within these fields are widely understood to drive the defining features of turbulent flows: multiscale energy transfer, intermittency, and the dissipative anomaly, i.e. the persistence of finite mean energy dissipation in the limit of vanishing viscosity \citep{frisch-1996,vassilicos-2015,mollo-christensen-1973,mukherjee-etal-2024,buaria-pumir-2026}. A fundamental question in fluid mechanics is identifying precisely which kinematic ingredients of the nonlinear interactions sustain these hallmarks \citep{ishihara-gotoh-kaneda-2009,eyink-2024}. 

Central to this inquiry is the vortex stretching mechanism, $\omega_j \partial_j u_i$, which arises from the nonlinear term in the vorticity equation,
\begin{equation}
\partial_t \omega_i + u_j \partial_j \omega_i =
\omega_j \partial_j u_i + Re^{-1} \partial_j^2 \omega_i,
\label{eq:vor-eq}
\end{equation}
where $u_i$ and $\omega_i$ denote the components of the velocity and vorticity fields, respectively, $Re$ is the Reynolds number and the Einstein summation convention is implied for repeated indices. This term encapsulates the action of local velocity gradients on the vorticity field and can be strictly decomposed into two distinct physical processes. The first is vorticity amplification (VA), $\omega_j \partial_j u_\ell \omega_\ell \omega_i \omega^{-2}$, which actively produces enstrophy by stretching fluid elements. The second is vortex tilting (VT), $(\delta_{i \ell} - \omega_i \omega_\ell \omega^{-2}) \omega_j \partial_j u_\ell$, which governs the reorientation of vorticity vectors without directly altering their magnitude.

Vortex stretching has long been recognized as the primary engine of three-dimensional turbulence \citep{taylor-1938,betchov-1956,ashurst-etal-1987}. Historically, VA is associated with the generation of small scales \citep{tsinober-2000,davidson-2004}, the forward transfer of energy across the inertial range \citep{davidson-etal-2008,doan-etal-2018,johnson-2020,carbone-bragg-2020}, and the amplification of intense velocity gradients that form localized regions of extreme enstrophy production \citep{she-etal-1990,jimenez-2000,buaria-etal-2020,buaria-etal-2024}. By contrast, VT dictates the geometric reorganization, strain-vorticity alignment, and topological complexity of the vortex lines \citep{tsinober-2000,holzner-etal-2010}. 
Recent theoretical and numerical efforts have provided new insights by exploring modified Navier--Stokes equations in which the total vortex stretching (VA+VT) is artificially suppressed. These studies reveal striking departures from classical Kolmogorov phenomenology \citep{bos-2021,bos-2025}. Helicity and enstrophy emerge as inviscid invariants, while energy is not conserved in the inviscid limit. The resulting system exhibits a forward cascade of enstrophy akin to two-dimensional turbulence \citep{boffetta-ecke-2012}, with the energy spectrum exhibiting a $E(k) \sim k^{-3}$ scaling in sharp contrast with the $E(k) \sim k^{-5/3}$ Kolmogorov predictions. In the case of helical flows, an additional inverse helicity cascade occurs and the flow dynamics relaxes toward large-scale force-free states \citep{wu-bos-2021,wu-bos-2022}. However, because these modifications eliminate both mechanisms simultaneously, the distinct, individual contributions of VA and VT remain obscured. In particular, it is an open question to what extent the anomalous dissipation of energy and the manifestation of intermittency originate strictly from the amplification of vorticity magnitude, the geometric reorientation of vortex lines, or their non-trivial interplay.

In the present work, we address this gap by isolating the respective contributions of VA and VT in homogeneous isotropic turbulence. Using direct numerical simulations, we investigate a modified Navier--Stokes system in which VA is selectively suppressed while VT is explicitly retained. We demonstrate that removing VA critically weakens small-scale velocity gradients, suppresses extreme fluctuation events, and entirely eliminates the classical energy dissipative anomaly. Remarkably, however, anomalous scaling and multifractal intermittency are shown to persist. Furthermore, we show that in this VA-suppressed limit, enstrophy assumes the role of the relevant inviscid invariant and exhibits its own anomalous dissipation.

The remainder of this paper is organized as follows. In \S\ref{sec:methods}, we detail the numerical methodology and the formulation of the modified Navier--Stokes equations. In \S\ref{sec:results}, we present our results regarding the suppression of the dissipative anomaly and the persistence of intermittency. Finally, conclusions and physical implications are discussed in \S\ref{sec:conclusions}.

\section{Methods}
\label{sec:methods}

\subsection{The Modified Equations}

To elucidate the distinct roles of vorticity amplification (VA) and vortex tilting (VT) in shaping turbulence dynamics, we perform direct numerical simulations (DNS) of homogeneous isotropic turbulence in an incompressible Newtonian fluid. We manipulate the Navier--Stokes equations by introducing a volume forcing that selectively suppresses specific components of vortex stretching within the nonlinear dynamics. The governing equations read
\begin{equation}
\partial_t u_i + u_j \partial_j u_i = - \partial_i p + Re^{-1} \partial_j^2 u_i + f_i + \gamma g_i, \quad \partial_i u_i = 0,
\label{eq:ns}
\end{equation}
where $p$ is the pressure, $f_i$ is a large-scale forcing used to sustain statistically stationary turbulence, and $g_i$ is an additional scale-dependent volume force constructed to remove the targeted component of the vortex stretching. The parameter $\gamma \in [0,1]$ controls the strength of the modification: $\gamma = 0$ recovers the classical Navier--Stokes dynamics, while $\gamma = 1$ corresponds to the complete removal of the targeted contribution.

The scale-dependent forcing $g_i$ is constructed from the vorticity evolution equation \eqref{eq:vor-eq} by explicitly isolating the targeted mechanisms and subtracting their contribution from the momentum equation. Specifically, we decompose the vortex stretching term $\omega_j \partial_j u_i \equiv \omega_j s_{ij}$ (where $s_{ij} = (\partial_j u_i + \partial_i u_j )/2$ is the symmetric rate-of-strain tensor) into two orthogonal components, VA and VT, by projecting it parallel and perpendicular to the local vorticity vector:
\begin{equation}
\omega_j \partial_j u_i = \underbrace{\omega_j \partial_j u_\ell \, \omega_\ell \omega_i \omega^{-2}}_{\text{VA}_i} + \underbrace{(\delta_{i\ell} - \omega_\ell \omega_i \omega^{-2}) \omega_j \partial_j u_\ell}_{\text{VT}_i},
\end{equation}
where $\omega^2 = \omega_i \omega_i$. 
This geometric decomposition ensures that the inner product of the VA term with $\omega_i$ yields exactly the enstrophy production term ($\text{VA}_i \omega_i = \omega_i \omega_j s_{ij}$), demonstrating that VA exclusively modulates the overall magnitude of the vorticity vector. Conversely, the VT term is identically orthogonal to the vorticity ($\text{VT}_i \omega_i = 0$). While physically originating from $s_{ij}$, this orthogonal component does not alter the vorticity vector's magnitude, but solely drives the geometrical reorientation (tilting) of the vortex lines.

Substituting this decomposition into the vorticity equation and applying the curl operator $\epsilon_{ink} \partial_n$ yields
\begin{align}
\epsilon_{ink} \partial_n \partial_t \omega_k + \epsilon_{ink} \partial_n (u_j \partial_j \omega_k) &= \epsilon_{ink} \partial_n \left[ \omega_j \partial_j u_\ell \omega_\ell \omega_k \omega^{-2} \right] \nonumber \\
&\quad + \epsilon_{ink} \partial_n \left[ (\delta_{k\ell} - \omega_\ell \omega_k \omega^{-2}) \omega_j \partial_j u_\ell \right] + Re^{-1} \epsilon_{ink} \partial_n \partial_j^2 \omega_k.
\end{align}
Using the vector identity $\epsilon_{ink} \partial_n \epsilon_{k\ell s} \partial_\ell u_s = \partial_i \partial_j u_j - \partial_j^2 u_i$, alongside the incompressibility condition $\partial_j u_j = 0$, we recast the equation in terms of the velocity field:
\begin{align}
\partial_t \partial_j^2 u_i - \epsilon_{ink} \partial_n (u_j \partial_j \omega_k) &+ \epsilon_{ink} \partial_n \left[ \omega_j \partial_j u_\ell \omega_\ell \omega_k \omega^{-2} \right] \nonumber \\
&+ \epsilon_{ink} \partial_n \left[ (\delta_{k\ell} - \omega_\ell \omega_k \omega^{-2}) \omega_j \partial_j u_\ell \right] = Re^{-1} \partial_j^2 \partial_\ell^2 u_i.
\end{align}
Applying the inverse Laplacian $(\partial_\ell^2)^{-1}$, which is well-defined in a periodic domain for zero-mean fields, we recover the momentum equation:
\begin{align}
\partial_t u_i - (\partial_\ell^2)^{-1} \epsilon_{ink} \partial_n (u_j \partial_j \omega_k) &+ (\partial_\ell^2)^{-1} \epsilon_{ink} \partial_n \left[ \omega_j \partial_j u_\ell \omega_\ell \omega_k \omega^{-2} \right] \nonumber \\
&+ (\partial_\ell^2)^{-1} \epsilon_{ink} \partial_n \left[ (\delta_{k\ell} - \omega_\ell \omega_k \omega^{-2}) \omega_j \partial_j u_\ell \right] = - \partial_i p + Re^{-1} \partial_j^2 u_i.
\end{align}
Note that the pressure gradient, $-\partial_i p$, is reintroduced to enforce incompressibility, as it identically vanishes under the curl operation ($\epsilon_{k\ell i} \partial_\ell \partial_i p = 0$).

This sequence formally demonstrates that the standard nonlinear advective term $u_j \partial_j u_i$ can be exactly partitioned into vorticity advection, VA and VT. The selective removal of a specific mechanism is therefore achieved by defining $g_i$ as the solution to a corresponding Poisson equation that counterbalances the targeted term. 
When the objective is to selectively suppress VA, the forcing is defined by
\begin{equation}
\partial_m^2 g_i = \epsilon_{ink} \partial_n \left( \omega_j \partial_j u_\ell \omega_\ell \omega_k \omega^{-2} \right).
\label{eq:forcing_VA}
\end{equation}
In this case energy is no longer conserved in the inviscid limit, and for $\gamma=1$ enstrophy becomes the sole relevant inviscid invariant (see \S\ref{sec:inv}). 
Alternatively, for the suppression of VT only, we set
\begin{equation}
\partial_m^2 g_i = \epsilon_{ink} \partial_n \left[ \left( \delta_{k\ell} - \omega_\ell \omega_k \omega^{-2} \right) \omega_j \partial_j u_\ell \right],
\label{eq:forcing_VT}
\end{equation}
and for the complete suppression of all vortex stretching (VA+VT), the forcing simplifies to
\begin{equation}
\partial_m^2 g_i = \epsilon_{ink} \partial_n (\omega_j \partial_j u_k).
\label{eq:forcing_VAVT}
\end{equation}

\subsection{Numerical Method and Simulation Parameters}
\label{sec:numerical_details}

We solve the modified Navier--Stokes equations \eqref{eq:ns} in a triply periodic domain of size $L_x=L_y=L_z=2\pi$ using an in-house parallelized finite-difference solver. The spatial derivatives are evaluated using a second-order central discretization scheme, and the system is advanced in time using a third-order Runge--Kutta scheme. Incompressibility is strictly enforced at each substep through a fast Poisson solver. The time step is dynamically adjusted to maintain the Courant--Friedrichs--Lewy (CFL) number below unity. Statistical quantities are collected only after the flow reaches a statistically stationary state and are averaged over sufficiently long integration times to ensure convergence.

Turbulence is primarily sustained by the Arnold--Beltrami--Childress (ABC) forcing \citep{podvigina-pouquet-1994}, defined as
\begin{equation}
\bm{f}_{\text{\tiny ABC}} =
\begin{pmatrix}
A \sin\left( k_f z \right) + C \cos \left( k_f y \right) \\
B \sin\left( k_f x \right) + A \cos \left( k_f z \right) \\
C \sin\left( k_f y \right) + B \cos \left( k_f x \right)
\end{pmatrix},
\label{eq:forcing_abc}
\end{equation}
where $k_f$ denotes the forcing wavenumber, fixed to $k_f=4$ in all simulations, with amplitudes $A=B=C=1$. Because the ABC forcing satisfies the Beltrami property, it injects both energy and helicity at the forcing scale. This permits the development of an inverse helicity cascade when the full vortex stretching (VA+VT) is entirely suppressed, allowing the flow to relax toward a large-scale condensate \citep{wu-bos-2021,wu-bos-2022}.
Because the volume forcing does not impose a constant energy injection rate, the resulting Taylor-scale Reynolds number, $Re_\lambda = u_{\mathrm{rms}}\lambda/\nu$ (where $u_{\mathrm{rms}}$ is the root-mean-square velocity and $\lambda$ is the Taylor microscale), varies as a function of the suppression parameter $\gamma$. 

To systematically assess Reynolds-number effects, two primary sets of simulations are performed. In the reference Navier--Stokes limit ($\gamma=0$), the kinematic viscosity $\nu$ is adjusted to obtain baseline Reynolds numbers of $Re_\lambda \approx 65$ and $Re_\lambda \approx 120$. For each set, $\gamma$ is varied continuously over the interval $[0,1]$. Unless otherwise specified, all results presented throughout the manuscript correspond to the higher-Reynolds-number configurations. For $\gamma=1$, additional simulations are performed up to $Re_\lambda \approx 4500$ to properly investigate the influence of VA on the dissipative anomaly and intermittency.

For comparative analysis, two alternative configurations are also simulated using viscosities matched to these baseline cases. In the first, the full vortex stretching term (VA+VT, $\gamma=1$) is removed. In the second, only VT is suppressed, leaving VA intact. Suppressing VT promotes the artificial alignment of vortex lines, thereby strongly enhancing VA and rapidly generating small-scale structures. Consequently, a small suppression parameter ($\gamma=0.05$) is employed in the VT-suppressed case, as larger values would require prohibitively high spatial resolution.

Adequate spatial resolution is strictly enforced across all parameter regimes using uniform grids of $512^3$ and $1024^3$ points. For the standard Navier--Stokes simulations ($\gamma=0$), the spatial resolution is verified against the classical Kolmogorov length scale, $\eta=\nu^{3/4}\langle \varepsilon \rangle^{-1/4}$, ensuring $\eta/\Delta x \sim \mathcal{O}(1)$. However, as $\gamma \to 1$, the forward transfer of kinetic energy is suppressed, and the relevant dissipative length scale becomes governed by enstrophy, scaling as $\eta_\omega=\nu^{1/2}\langle \varepsilon_\omega \rangle^{-1/6}$ \citep{bos-2021}. Here, $\varepsilon$ and $\varepsilon_\omega$ denote the kinetic-energy and enstrophy dissipation rates, respectively, and $\langle \cdot \rangle$ indicates the averaging operator. At higher Reynolds numbers, grid resolutions of up to $1024^3$ are utilized to strictly maintain $\eta_\omega/\Delta x > 1$ through the maximum $Re_\lambda \approx 4500$. Comprehensive grid-independence tests and a detailed breakdown of the resolution parameters are provided in the Appendix \ref{sec:appendix}.

Finally, to investigate the dependence of the observed dynamics on the large-scale injection mechanism, an additional dataset is generated using a non-helical (NH) forcing scheme, given by
\begin{equation}
\bm{f}_{\text{ \tiny NH}} =
\begin{pmatrix}
A \cos\left( k_f z \right) \cos \left( k_f y \right) \\
B \cos\left( k_f x \right) \cos \left( k_f z \right) \\
C \cos\left( k_f y \right) \cos \left( k_f x \right)
\end{pmatrix}.
\label{eq:forcing_nh}
\end{equation}
For the non-helical simulations, the forcing is applied at the wavenumber $k_f = 1$, with the amplitude coefficients set to $A=B=C=1$. This specific dataset isolates the case of full VA removal ($\gamma=1$) and comprises seven simulations performed on a $512^3$ grid, with Reynolds numbers ranging between $Re_\lambda \in [160, 2600]$. For all of these cases, a strict resolution criterion of $\eta_\omega/\Delta x > 1.2$ is maintained (see \S\ref{sec:appendix}).

\section{Results}
\label{sec:results}

\subsection{Flow Visualisations, Spectra and Budgets}

\begin{figure}
\centering
\begin{tikzpicture}
  \node at (-6,0)  {\includegraphics[trim={550 0 550 0},clip,width=0.32\textwidth]{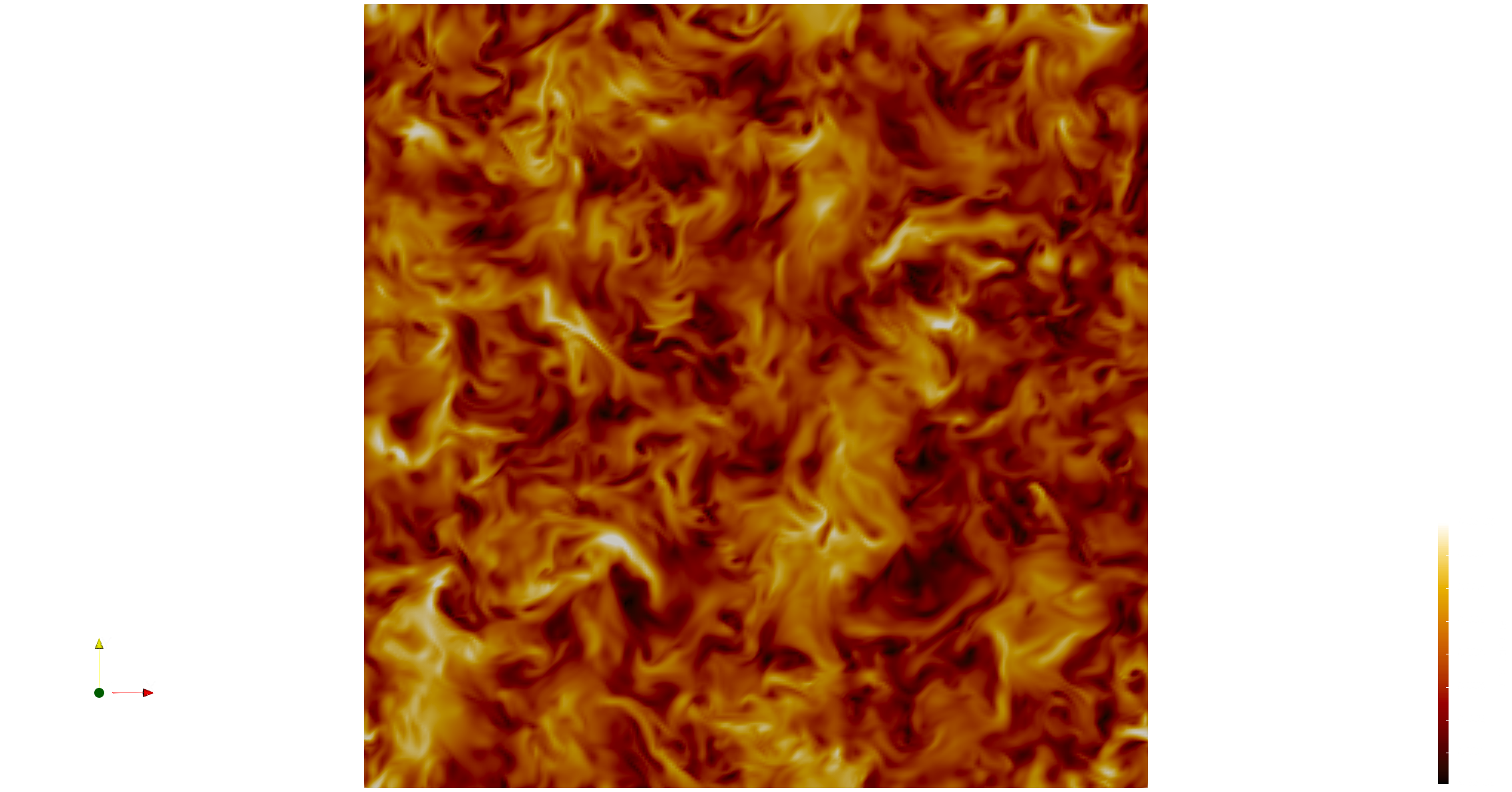}};
  \node at ( 0,0)  {\includegraphics[trim={550 0 550 0},clip,width=0.32\textwidth]{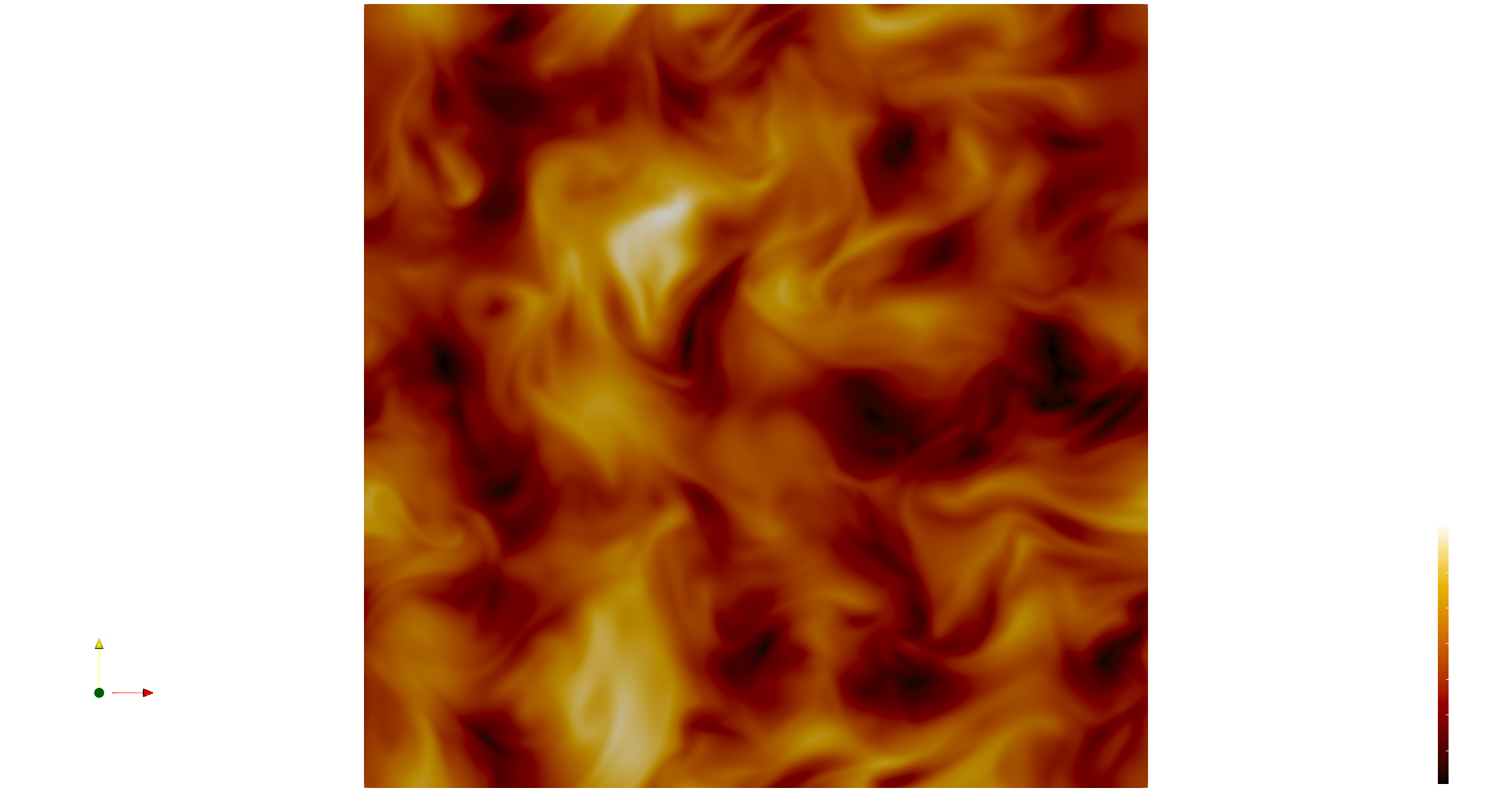}};
  \node at ( 6,0)  {\includegraphics[trim={550 0 550 0},clip,width=0.32\textwidth]{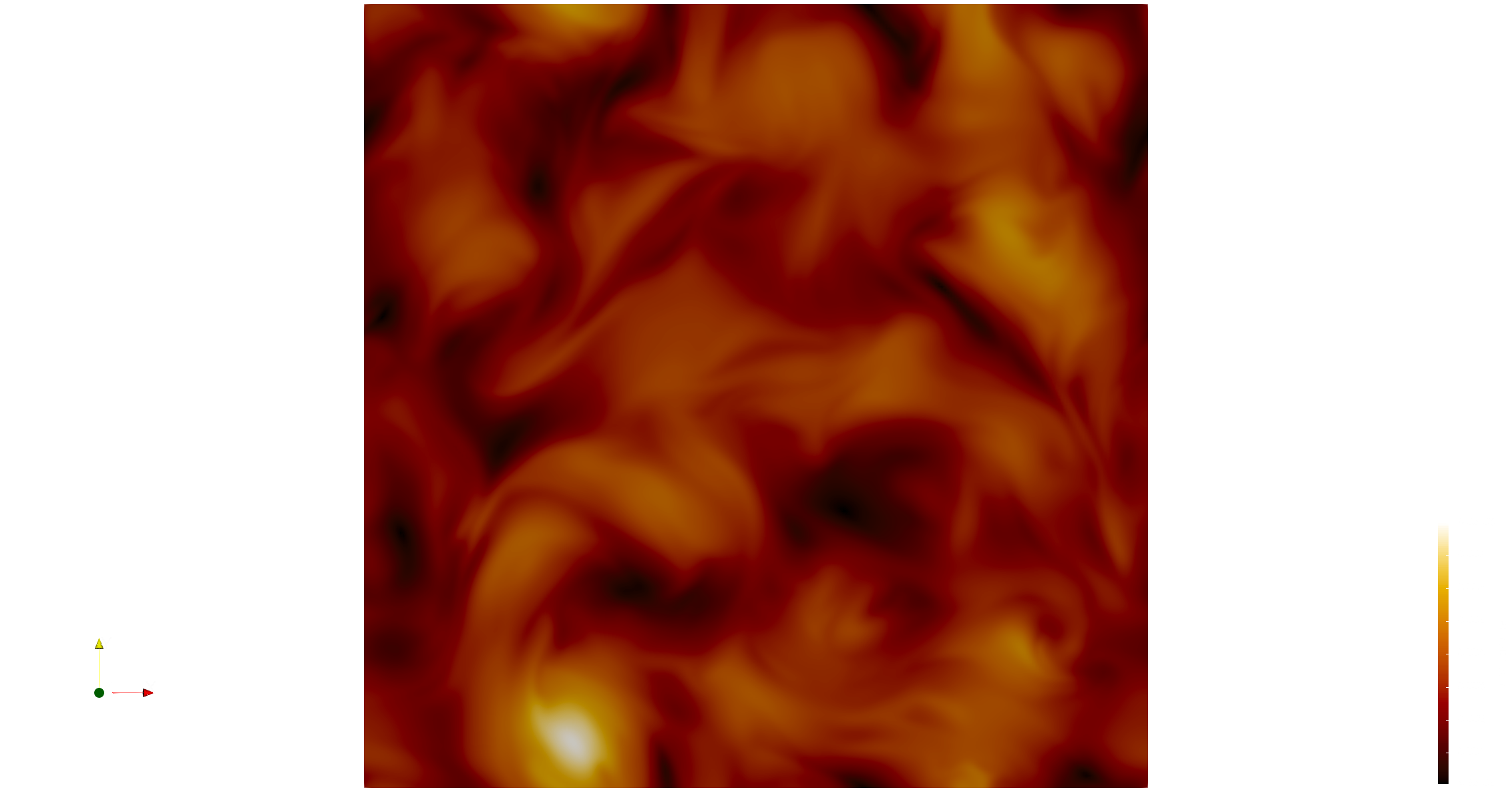}};
  \node at (-6,-6.2) {\includegraphics[trim={550 0 550 0},clip,width=0.32\textwidth]{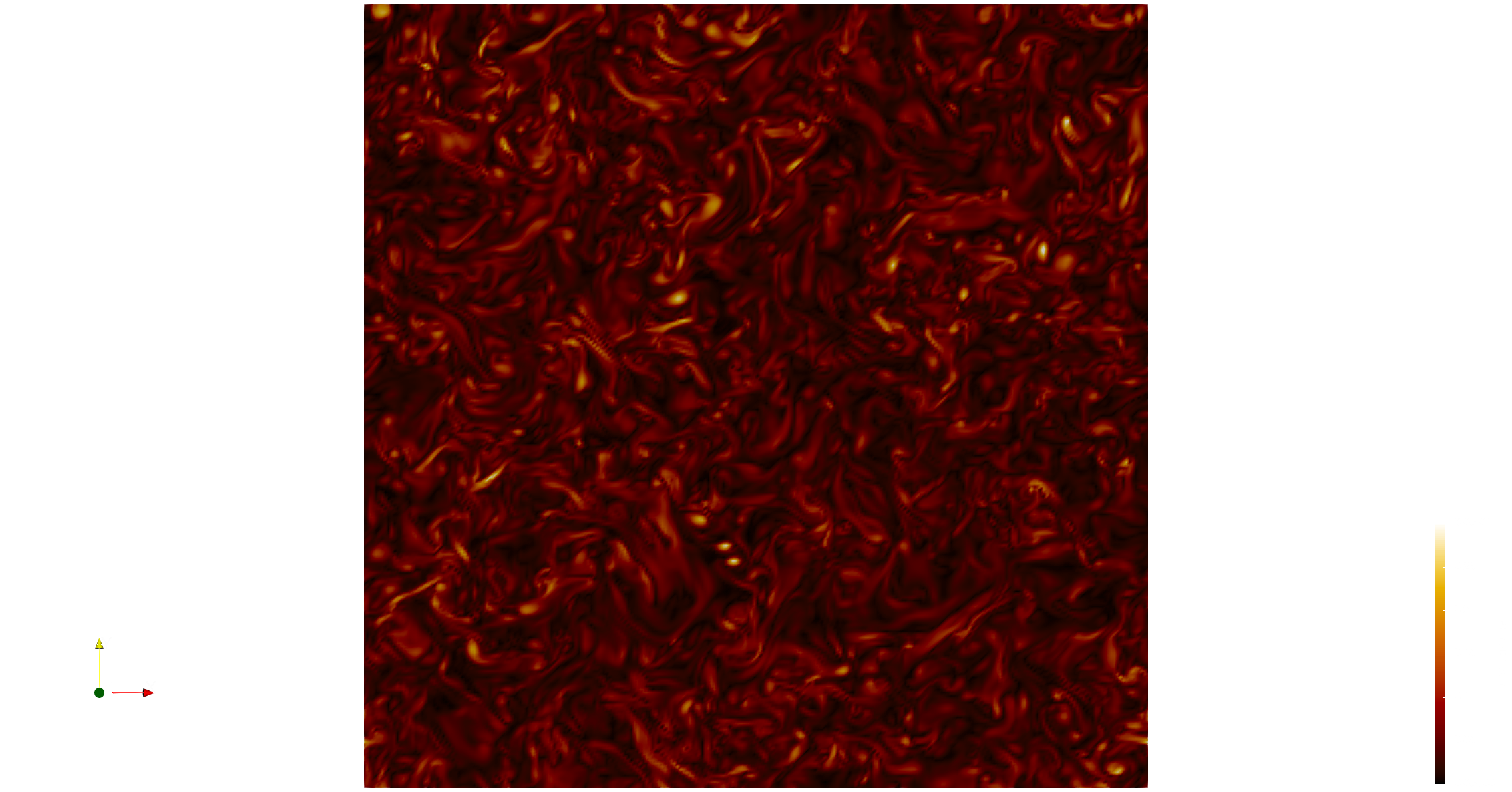}};
  \node at ( 0,-6.2) {\includegraphics[trim={550 0 550 0},clip,width=0.32\textwidth]{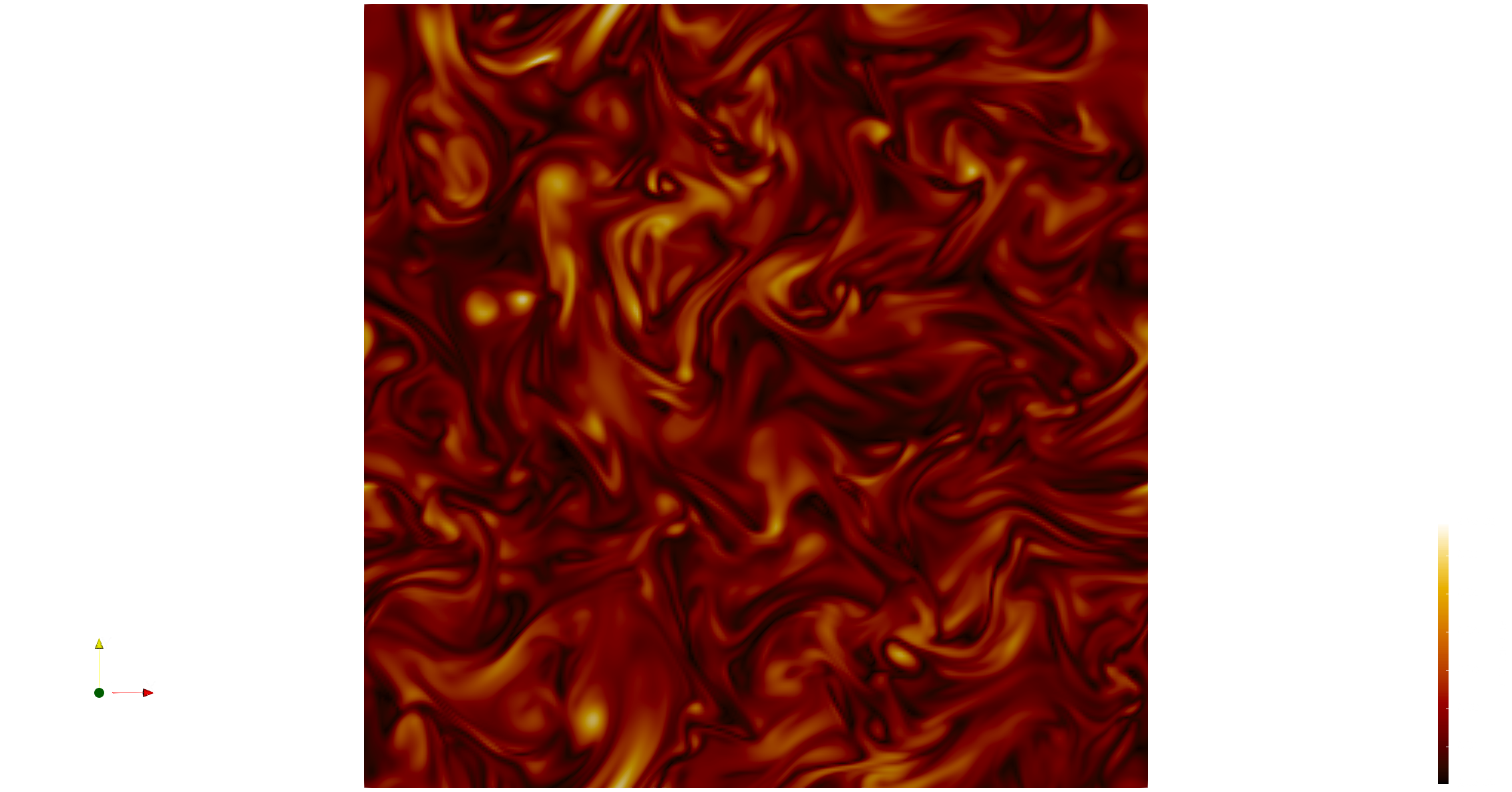}};
  \node at ( 6,-6.2) {\includegraphics[trim={550 0 550 0},clip,width=0.32\textwidth]{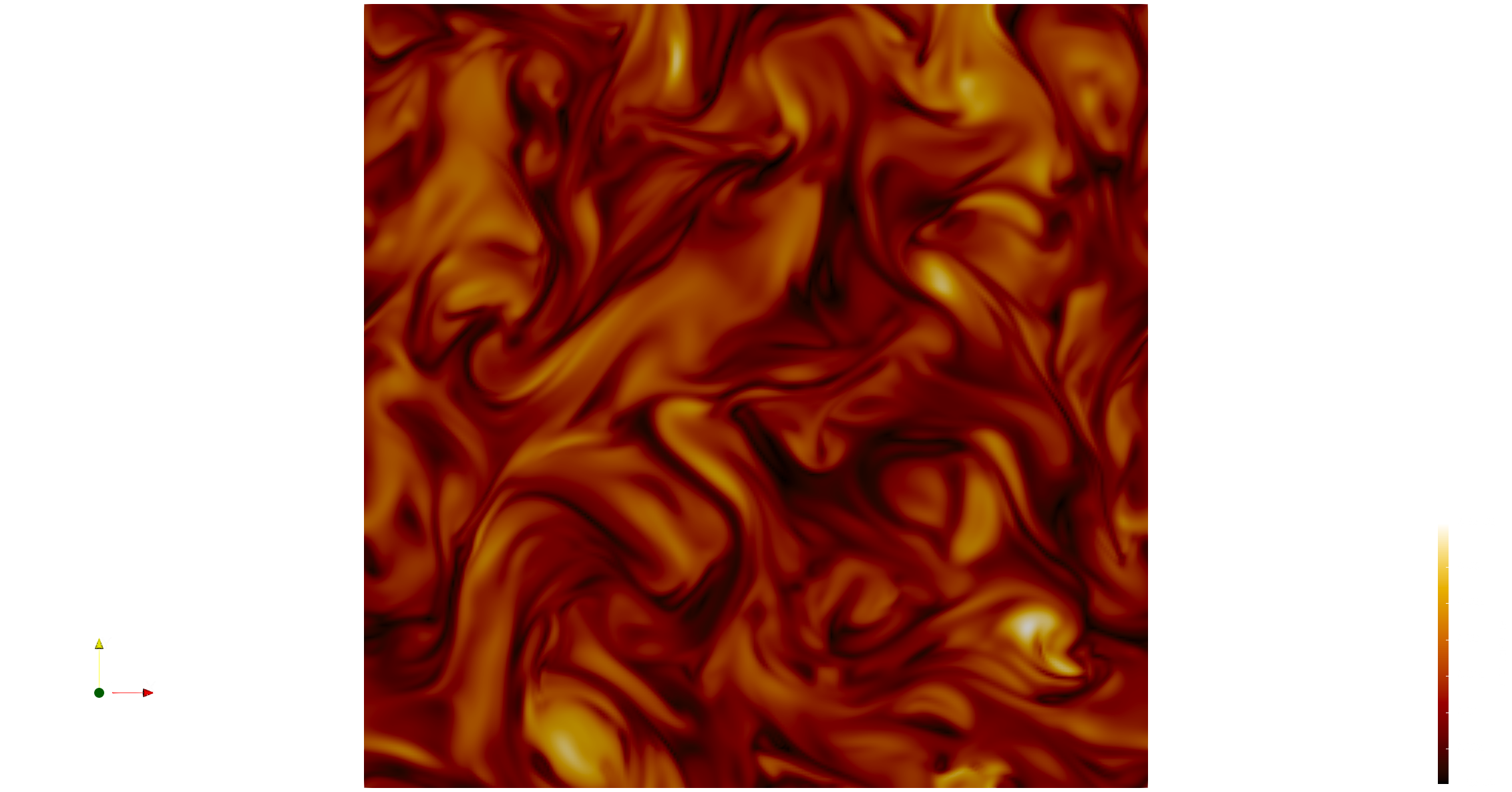}};
  \node at (-8.7,3.1) {(a)};
  \node at (-2.7,3.1) {(b)};
  \node at ( 3.3,3.1) {(c)};
  \node at (-8.7,-3.1){(d)};
  \node at (-2.7,-3.1){(e)};
  \node at ( 3.3,-3.1){(f)};
\end{tikzpicture}
\caption{
Effect of selective VA suppression on the spatial structure of the flow. Instantaneous two-dimensional slices of the kinetic-energy field, $q = u_i u_i$ (top row, a--c), and the enstrophy field, $\omega^2 = \omega_i \omega_i$ (bottom row, d--f). Columns correspond to increasing levels of suppression: standard Navier--Stokes turbulence ($\gamma=0$; a, d), partial suppression ($\gamma=0.5$; b, e), and complete suppression ($\gamma=1$; c, f). As $\gamma \to 1$, intense fine-scale filamentary structures vanish, replaced by diffuse, large-scale coherent patches. All fields are extracted from ABC-forced simulations, with the baseline $\gamma=0$ case corresponding to $Re_\lambda \approx 120$.
}
\label{fig:snap}
\end{figure}
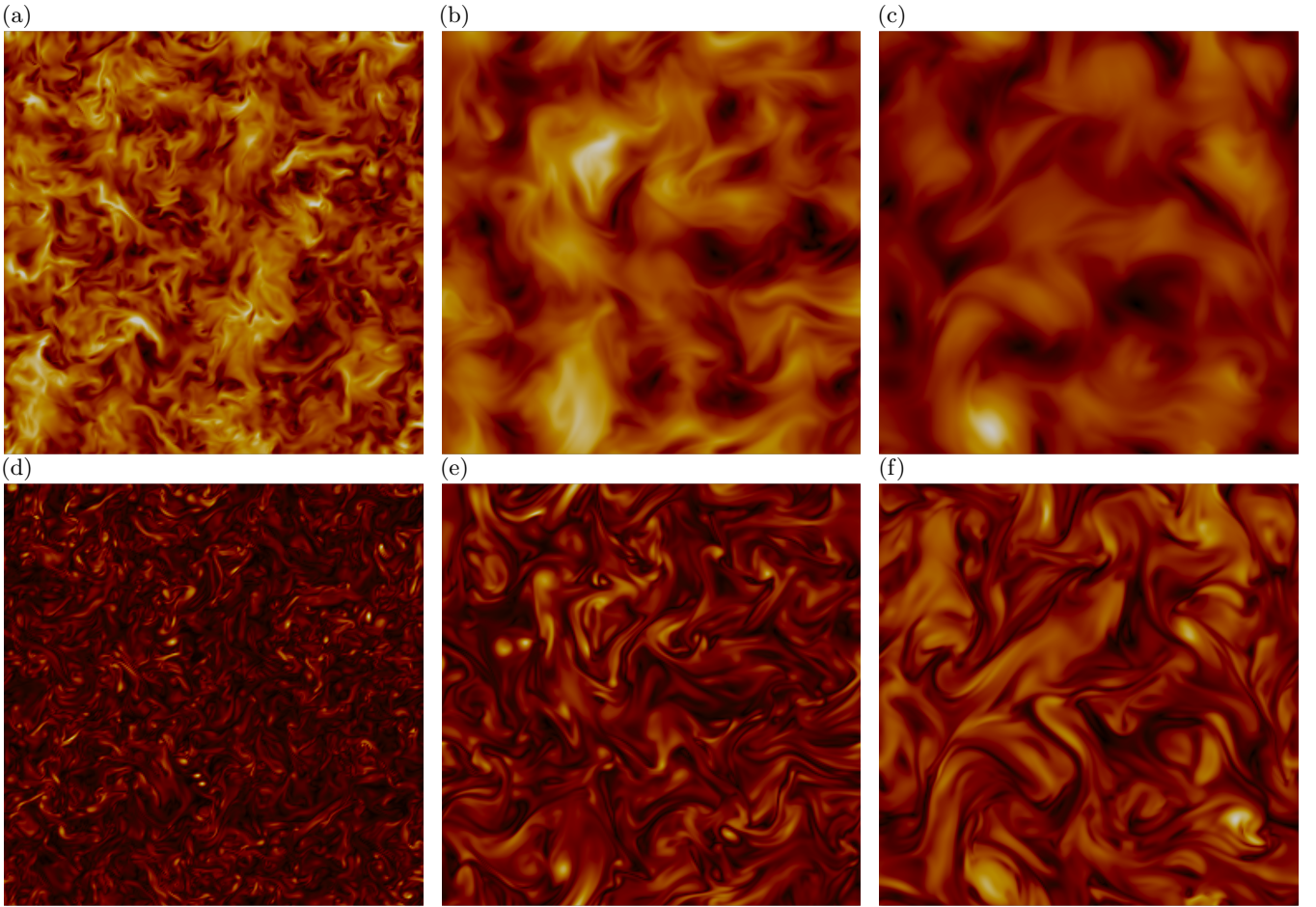

We begin by examining the qualitative impact of selectively suppressing VA on the spatial structure of the turbulent flow. Fig.~\ref{fig:snap}(a--f) presents instantaneous visualizations of the kinetic energy, $q = u_i u_i$, and enstrophy, $\omega^2 = \omega_i \omega_i$, fields for varying degrees of VA suppression. In the classical Navier--Stokes regime ($\gamma=0$, Figs.~\ref{fig:snap}a and \ref{fig:snap}d), the flow exhibits the familiar hallmarks of three-dimensional turbulence: energy is broadly distributed across a wide range of scales, and enstrophy is heavily concentrated in intense, fine-scale, filamentary vortex structures. 
As VA is progressively suppressed (e.g., $\gamma=0.5$, Figs.~\ref{fig:snap}b and \ref{fig:snap}e), the turbulent fields become noticeably smoother, reflecting a severe depletion of intense small-scale velocity gradients. In the limit of complete VA suppression ($\gamma=1$, Figs.~\ref{fig:snap}c and \ref{fig:snap}f), the fine-scale filamentary structures vanish entirely. Instead, the enstrophy organizes into diffuse, large-scale patches. Remarkably, despite this stark reduction in velocity gradients, the flow does not simply decay into an unstructured state; it retains a highly coherent, large-scale spatial organization. This visual evidence clearly indicates that VT alone remains sufficient to sustain the topological complexity and coherent structural organization of the turbulent flow, even in the complete absence of the stretching mechanism that traditionally amplifies its magnitude. Notably, unlike modified systems where the total vortex stretching (VA+VT) is suppressed \citep{wu-bos-2021,wu-bos-2022}, we do not observe the formation of a large-scale condensate. This is because helicity is not an inviscid invariant when only VA is selectively removed (see \S\ref{sec:inv}), precluding the inverse cascade mechanism responsible for condensate formation.

\begin{figure}
\centering
\begin{tikzpicture}
  \node at ( 0,-6) {\includegraphics[trim={0 15 0 0},clip,width=0.7\textwidth]{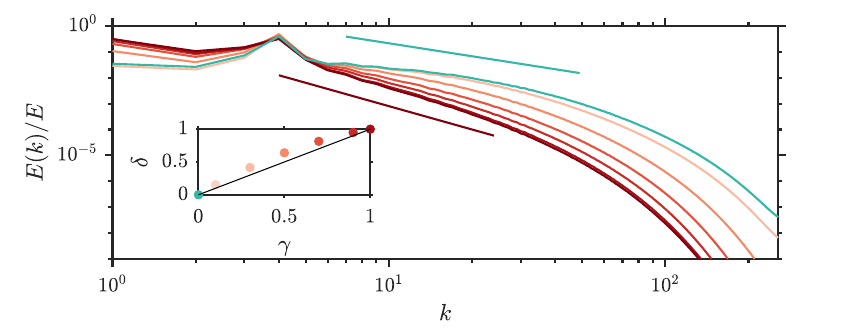}};
  \node at ( 0,-10.5) {\includegraphics[trim={0 5 0 0},clip,width=0.7\textwidth]{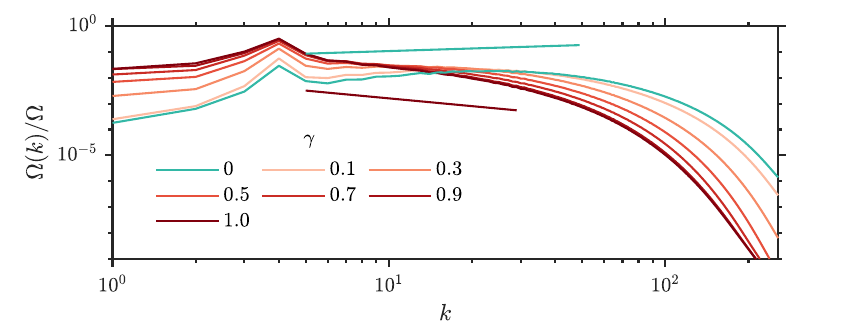}};
  \node at ( 0,-15){\includegraphics[trim={0  0 0 0},clip,width=0.7\textwidth]{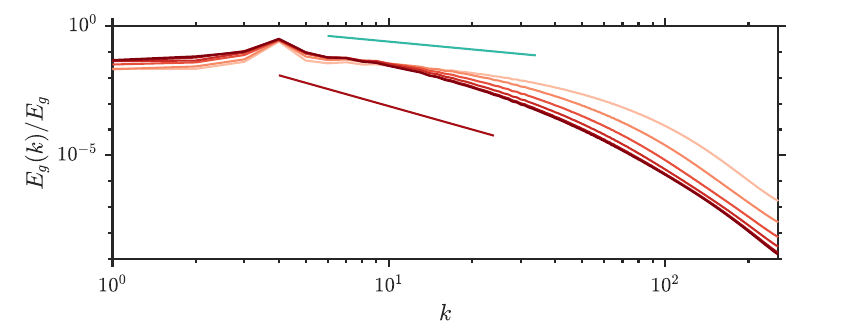}};
  \node at (-6,-4.2){(a)};
  \node at (-6,-8.55){(b)};
  \node at (-6,-13){(c)};
\end{tikzpicture}
\caption{Spectral signatures under progressive suppression of VA. (a) Kinetic energy $E(k)$ and (b) enstrophy $\Omega(k)$ spectra for varying suppression parameter $\gamma$. Solid green and red lines indicate classical Kolmogorov ($k^{-5/3}$, $k^{1/3}$) and fully suppressed asymptotic ($k^{-3}$, $k^{-1}$) scalings, respectively. Inset (a): Dependence of the exponent $\delta$ on $\gamma$, parameterizing the continuous transition of the intermediate scaling $E(k) \sim k^{-5/3-4\delta/3}$ from a standard energy ($\delta=0$) to an enstrophy cascade ($\delta=1$) \citep{bos-2025}. (c) Dynamically developed forcing spectrum, $E_g(k)$. As $\gamma \to 1$, the intermediate-range forcing spectrum slope steepens from approximately $k^{-1}$ (solid green line) to $k^{-3}$ (solid red line), consistent with renormalization-group predictions relating forcing exponents to energy scalings \citep{forster-etal-1977,mccomb-1990,falkovich-etal-2001}. All data correspond to the ABC forcing, with the $\gamma=0$ baseline at $Re_\lambda \approx 120$.}
\label{fig:spectra}
\end{figure}

The qualitative smoothing of the flow is quantitatively reflected in its spectral signatures, shown in Fig.~\ref{fig:spectra}. We compute the energy spectrum, $E(k)=\frac{1}{2} \int \Phi_{ii}(\bm{k})\delta(|\bm{k}|-k)\,\mathrm{d}^3\bm{k}$ (where $\Phi_{ij}$ is the velocity spectral tensor), alongside the corresponding enstrophy spectrum, $\Omega(k)\equiv k^2E(k)$. As $\gamma$ increases, both spectra exhibit an accumulation of energy at large scales ($k \lesssim k_f$) and a pronounced depletion at small scales ($k \gtrsim k_f$). Concurrently, the inertial-range scaling progressively departs from the classical Kolmogorov predictions of $E(k) \sim \langle \varepsilon \rangle^{2/3}k^{-5/3}$ and $\Omega(k) \sim \langle \varepsilon \rangle^{2/3}k^{1/3}$ \citep{frisch-1996}.

The most extreme departure occurs in the limit of complete VA suppression ($\gamma=1$). Under this condition, energy is no longer conserved in the inviscid dynamics; instead, enstrophy emerges as the sole relevant inviscid invariant (see \S\ref{sec:inv}). The dynamics are therefore characterized by a forward enstrophy cascade (as demonstrated below), and the flow statistics in the intermediate range of scales are completely determined by the enstrophy dissipation rate, $\varepsilon_\omega$. Dimensional analysis \citep{bos-2021} dictates that these spectra must scale as
\begin{equation}
E(k) = C \langle \varepsilon_\omega\rangle^{2/3}k^{-3},
\qquad
\Omega(k) = C \langle \varepsilon_\omega\rangle^{2/3}k^{-1},
\end{equation}
where $C$ is an $\mathcal{O}(1)$ dimensionless constant. This scaling is in excellent agreement with our numerical results. 

Interestingly, the applied scale-dependent forcing $g_i$ dynamically develops an intermediate-range spectrum close to $E_g(k)\sim k^{-3}$ as $\gamma \to 1$ (Fig.~\ref{fig:spectra}c). This behavior aligns tightly with renormalization-group (RG) predictions that relate the scaling exponents of the forcing to those of the energy spectrum \citep{forster-etal-1977,mccomb-1990,falkovich-etal-2001}. In RG approaches, turbulence driven at all scales by a self-similar Gaussian forcing with a spectrum $E_g(k)\sim k^{6-d-y}$ yields an energy spectrum scaling as $E(k)\sim k^{4-d-2y/3}$, where $d$ is the spatial dimension and $y$ is a scaling parameter. In three dimensions ($d=3$), setting $y=4$ recovers the classical Kolmogorov scaling $E(k)\sim k^{-5/3}$ driven by $E_g(k)\sim k^{-1}$. By contrast, substituting $y=6$ yields exactly $E(k)\sim k^{-3}$ driven by $E_g(k)\sim k^{-3}$, which is perfectly consistent with the scaling observed in our fully VA-suppressed limit.

For intermediate suppression values, $\gamma \in (0,1)$, the spectra exhibit robust power-law behaviors that transition smoothly between the classical $k^{-5/3}$ and the VA-free $k^{-3}$ scalings. This continuous transition can be contextualized using the recent theoretical framework of \citet{bos-2025}, who investigated intermediate statistics when the full vortex stretching term (VA+VT) is only partially suppressed. They proposed that neither energy nor enstrophy act as exact inviscid invariants in these intermediate states. Instead, the system conserves a mixed invariant termed ``mesostrophy'', defined as
\begin{equation}
\mu=\int k^{2\delta}E(k)\,\mathrm{d}k,
\end{equation}
where the parameter $\delta$ depends directly on the suppression strength \citep{bos-2025}. The limiting cases of $\delta=0$ and $\delta=1$ naturally recover the classical Navier--Stokes dynamics ($\gamma=0$) and the enstrophy-conserving dynamics ($\gamma=1$), respectively. Under mesostrophy conservation, the inertial-range statistics are governed by the dissipation rate of $\mu$, leading to the dimensional predictions
\begin{equation}
E(k)\sim k^{-5/3-4\delta/3},
\qquad
\Omega(k)\sim k^{1/3-4\delta/3}.
\end{equation}
Our spectral data clearly support the emergence of this intermediate power-law scaling regime. To quantitatively determine the exact $\delta$--$\gamma$ correspondence, we compute the exponent $\delta$ by enforcing the global conservation of the generalized invariant. Following \cite{bos-2025}, this is achieved by requiring the spectral integral of the nonlinear mesostrophy transfer rate to vanish, i.e., $\int_0^\infty T^\mu \mathrm{d}k = 0$. The resulting dependence of $\delta$ on $\gamma$ is reported in the inset of Fig.~\ref{fig:spectra}(a).

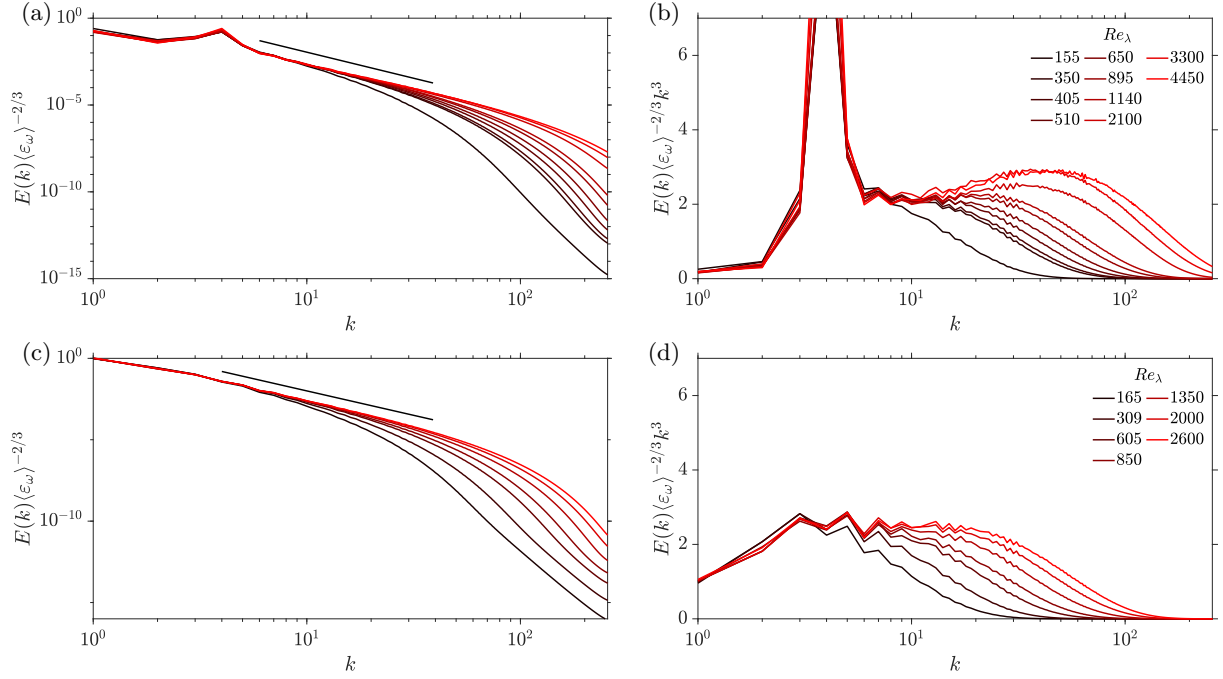
\begin{figure}
\centering
\begin{tikzpicture}
\node at (-4, 0.0) {\includegraphics[width=0.49\textwidth]{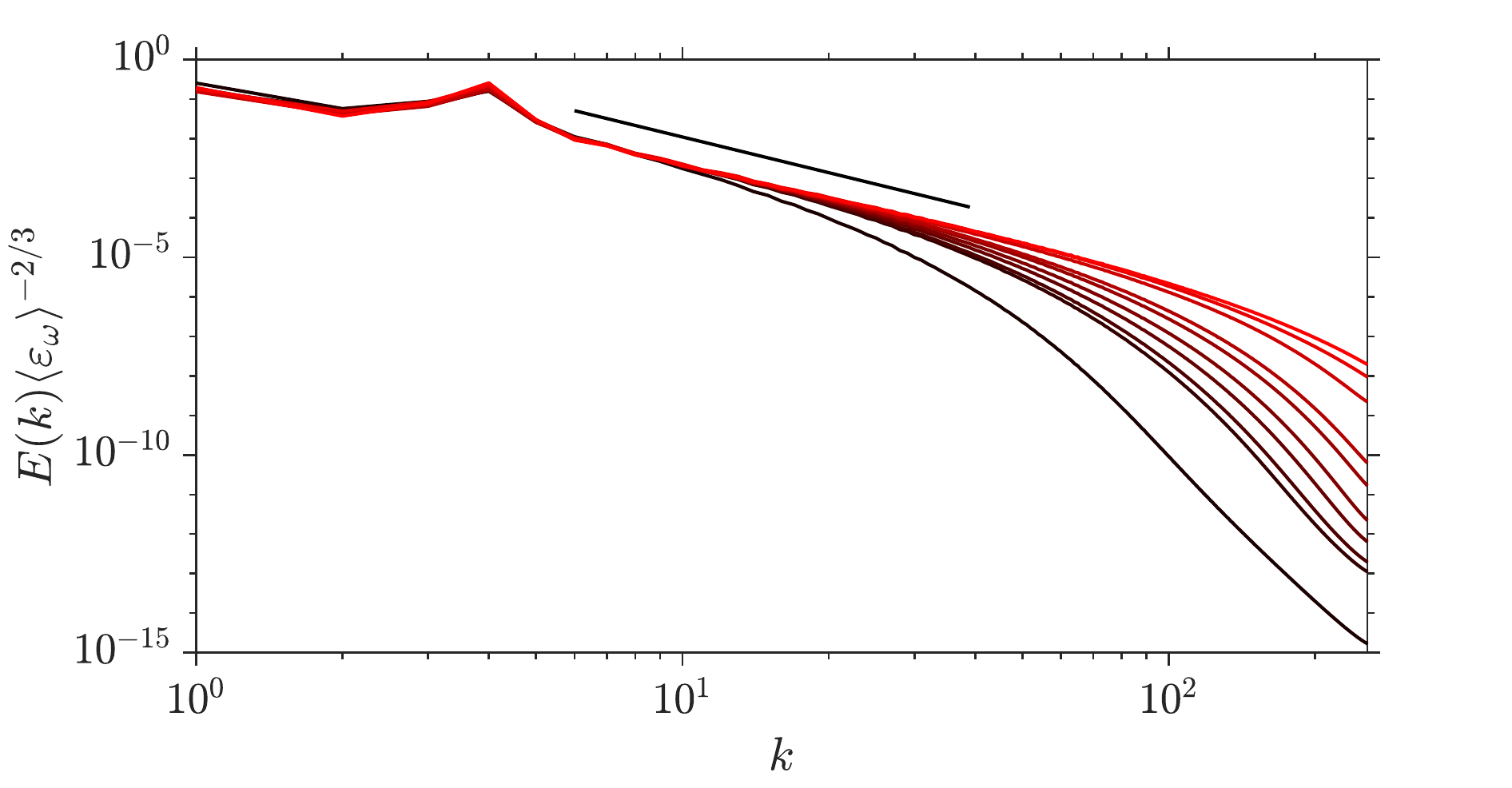}};
\node at ( 4, 0.0) {\includegraphics[width=0.49\textwidth]{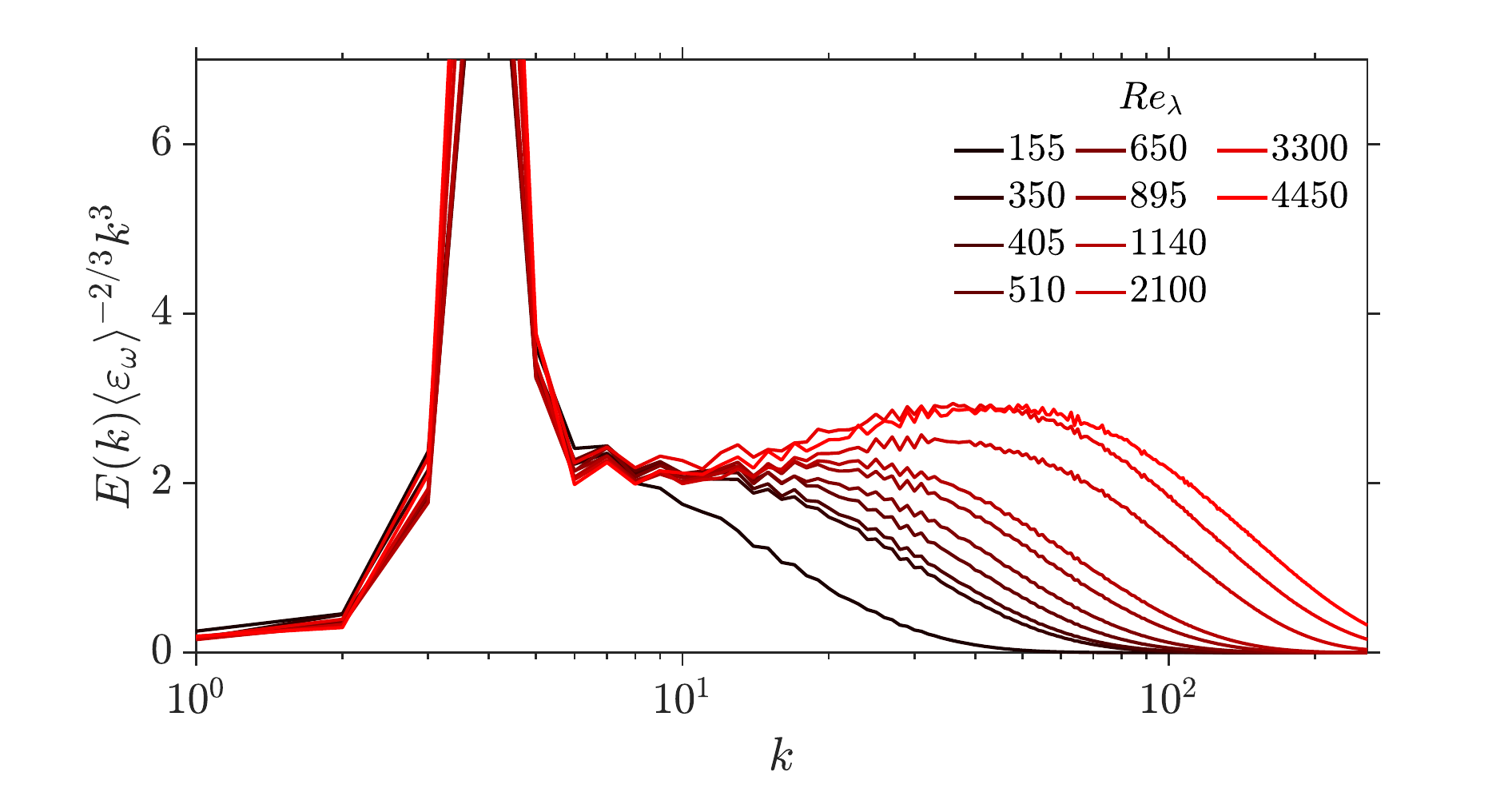}};
\node at (-4,-4.5) {\includegraphics[width=0.49\textwidth]{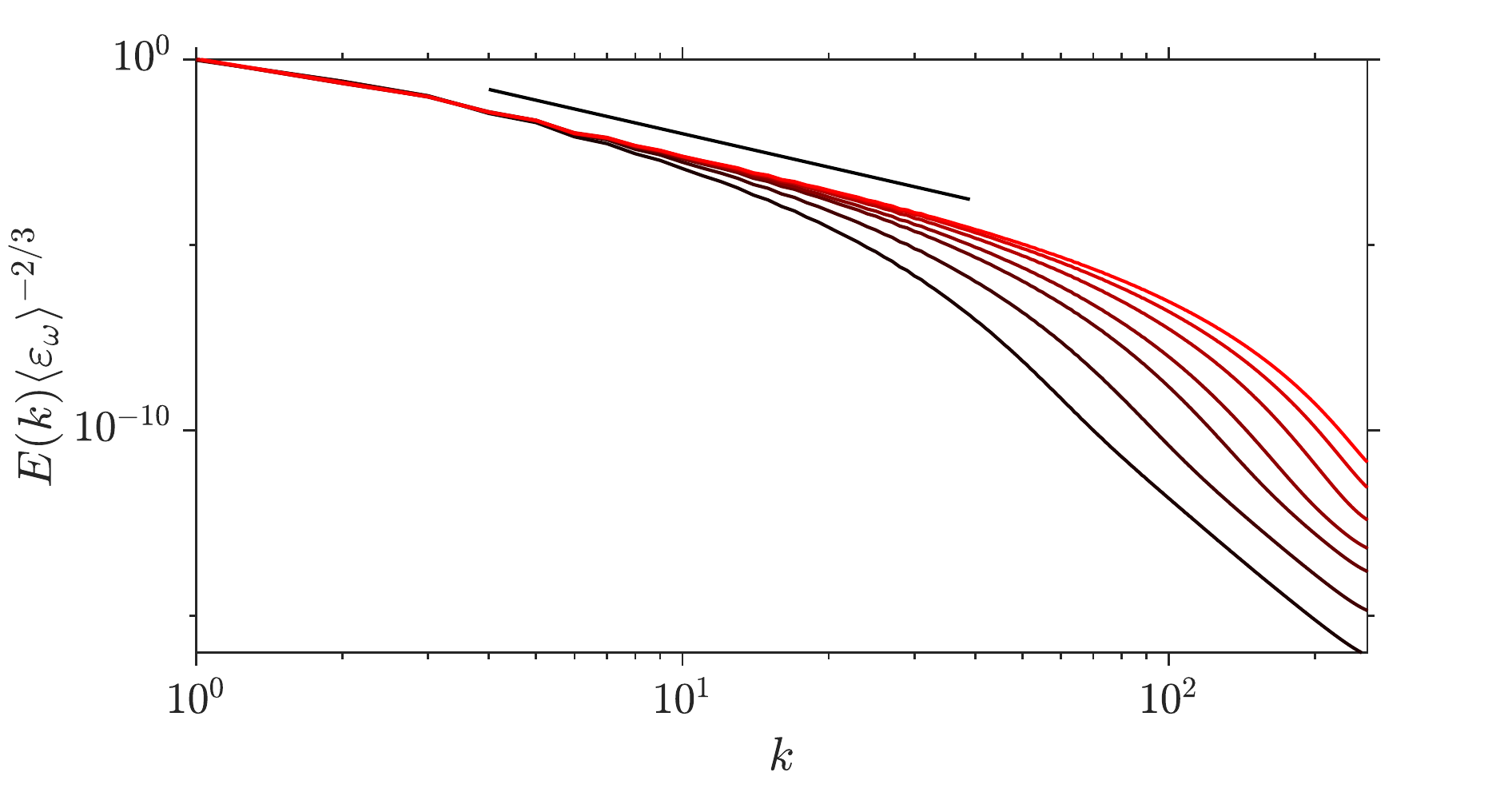}};
\node at ( 4,-4.5) {\includegraphics[width=0.49\textwidth]{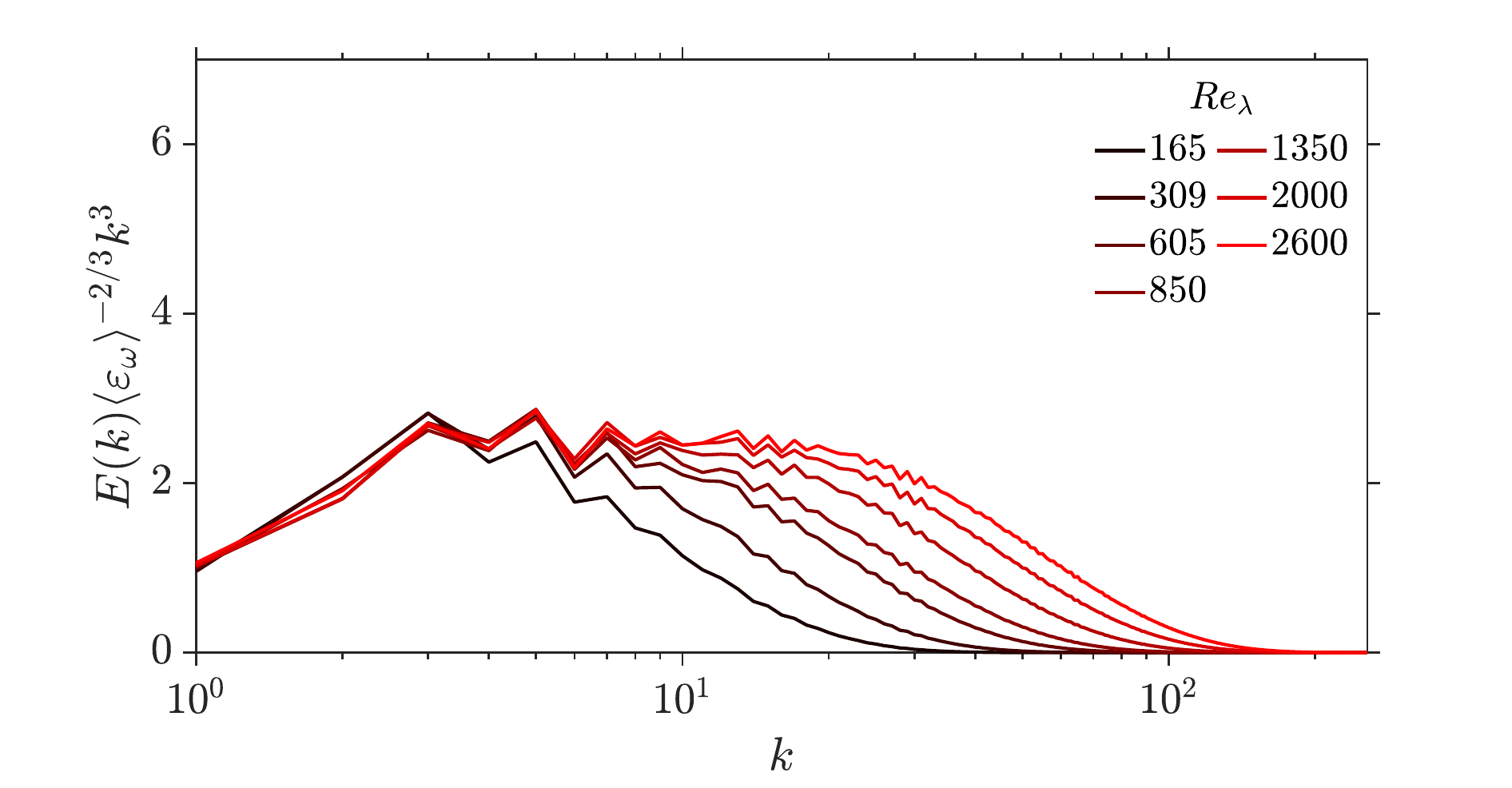}};
\node at (-8,2) {(a)};
\node at ( 0.3,2) {(b)};
\node at (-8,-2.5) {(c)};
\node at ( 0.3,-2.5) {(d)};
\end{tikzpicture}
\caption{Reynolds-number dependence and universality of energy spectra in the fully VA-suppressed limit ($\gamma=1$) for (a, b) helical ABC and (c, d) non-helical (NH) forcing. (a, c) Normalized energy spectra $E(k) \langle \varepsilon_\omega \rangle^{-2/3}$, with solid black lines indicating the theoretical $k^{-3}$ forward enstrophy cascade. (b, d) Compensated spectra $E(k)\langle \varepsilon_\omega \rangle^{-2/3} k^3$. Increasing $Re_\lambda$ yields an excellent collapse at intermediate scales, revealing an expanding $k^{-3}$ inertial range. The prominent plateaus in (b, d) confirm the universality of the enstrophy cascade, yielding a dimensionless constant $C \approx 2.1$--$2.4$ across both forcing schemes. Notably, ABC-driven flows exhibit a high-wavenumber bottleneck near the dissipation range (b), which is absent under NH forcing (d).}
\label{fig:spectra2}
\end{figure}

To verify the robustness of this $\gamma=1$ asymptotic state and to confirm its independence from the large-scale energy injection mechanism, Fig.~\ref{fig:spectra2} presents the Reynolds-number dependence of the energy spectra for both the ABC (top panels) and NH (bottom panels) forcing schemes. The distinct peak visible in the compensated spectra for the ABC cases is a direct signature of the specific energy injection scale ($k_f=4$). 
As $Re_\lambda$ increases, the normalized spectra $E(k) \langle \varepsilon_\omega \rangle^{-2/3}$ exhibit an excellent collapse at intermediate scales. The extent of the $k^{-3}$ scaling regime systematically widens, consistent with the development of an extensive enstrophy inertial range.
The compensated spectra, $E(k)\langle \varepsilon_\omega \rangle^{-2/3}k^3$ (Fig.~\ref{fig:spectra2}b,d), plateau precisely in this intermediate regime. This plateau provides a direct measurement of the universal dimensionless constant introduced above, yielding $C \approx 2.1$--$2.4$ across both forcing schemes. This robustly confirms the dimensional prediction for the enstrophy cascade. 
Interestingly, the high-Reynolds-number simulations driven by the ABC forcing reveal a distinct spectral bump near the onset of the dissipation range. This phenomenon, widely known as the bottleneck effect, has been extensively documented in classical Navier--Stokes turbulence \citep{falkovich-1994,lohse-mullergroeling-1995,dobler-etal-2003,meyers-meneveau-2008} and is generally attributed to the incomplete thermalization of triad interactions at large wavenumbers \citep{frisch-etal-2008}. Curiously, this bottleneck signature is absent in the non-helical (NH) forcing data, at least within the range of Reynolds numbers investigated here. While a detailed theoretical explanation of this forcing-dependent, high-$k$ behavior lies beyond the scope of the present work, its presence does not alter the universality of the intermediate inertial-range scaling or the estimation of the constant $C$.

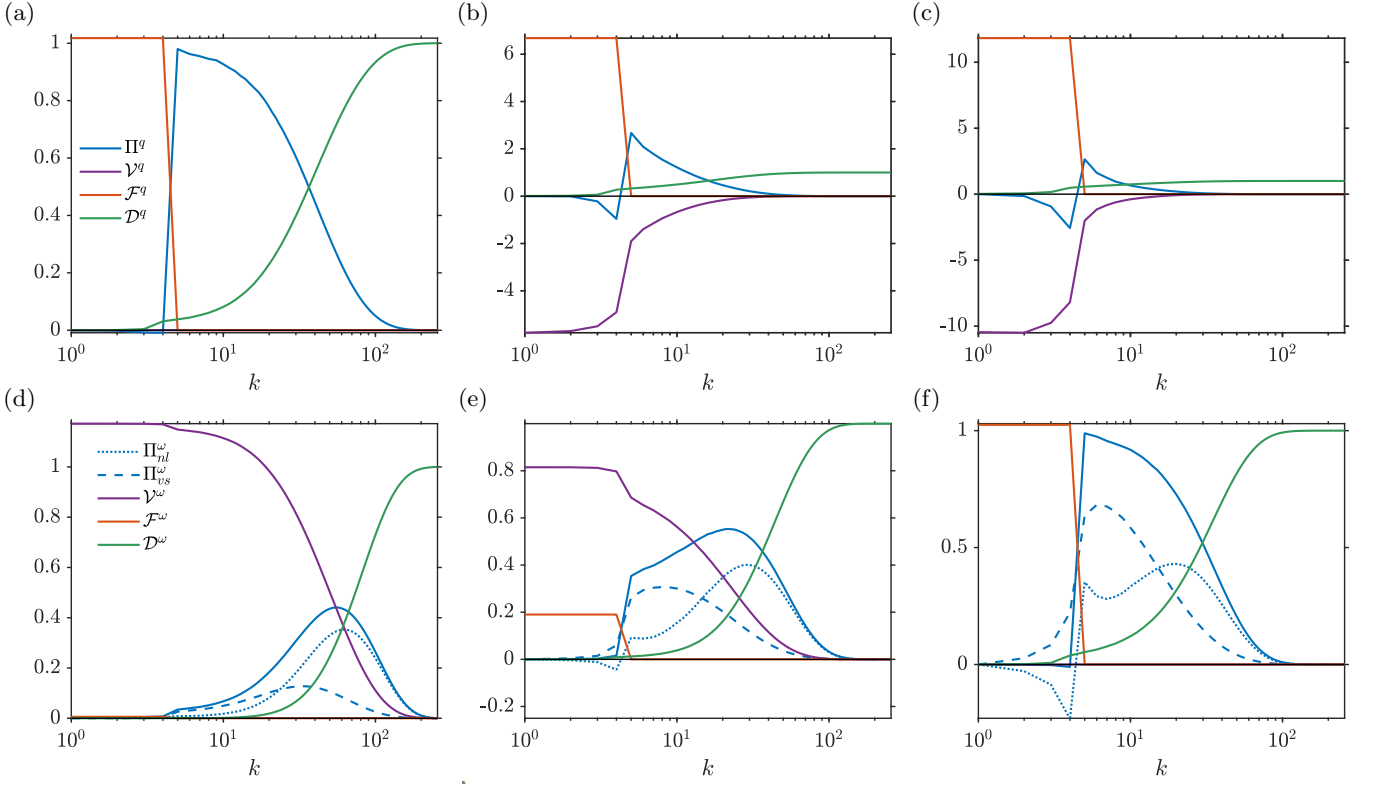
\begin{figure}
\centering
\begin{tikzpicture}
\node at (-6,0) {\includegraphics[width=0.35\textwidth]{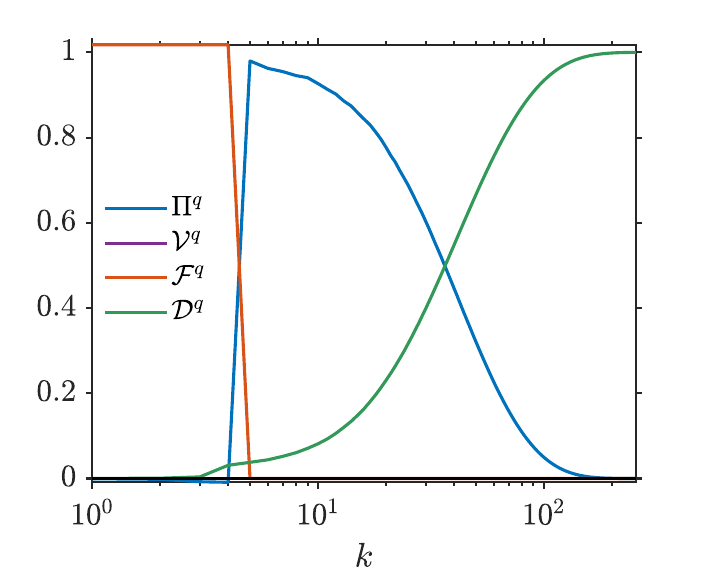}};
\node at ( 0,0) {\includegraphics[width=0.35\textwidth]{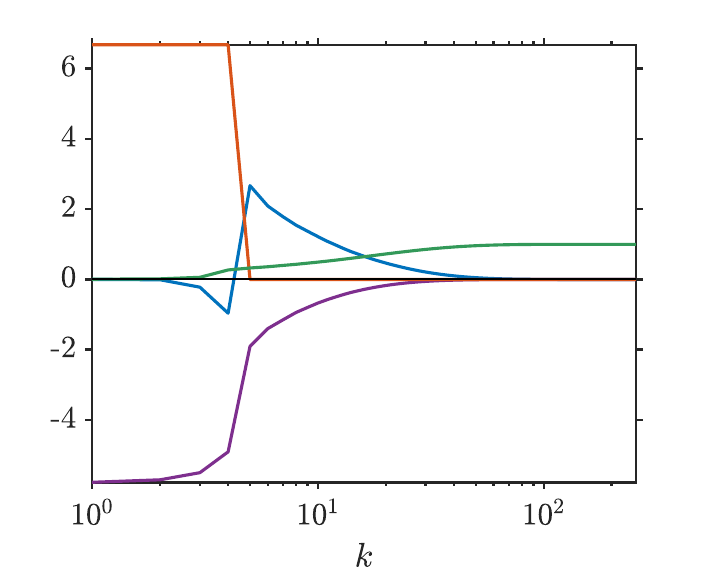}};
\node at ( 6,0) {\includegraphics[width=0.35\textwidth]{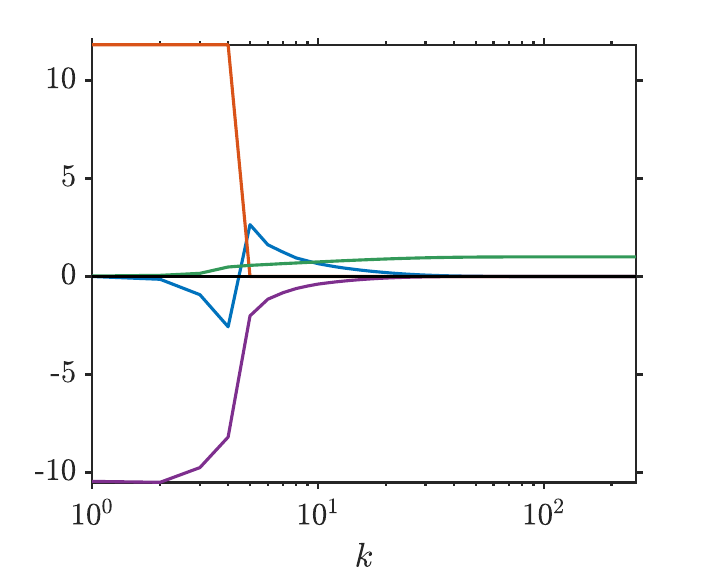}};
\node at (-6,-5.1){\includegraphics[width=0.35\textwidth]{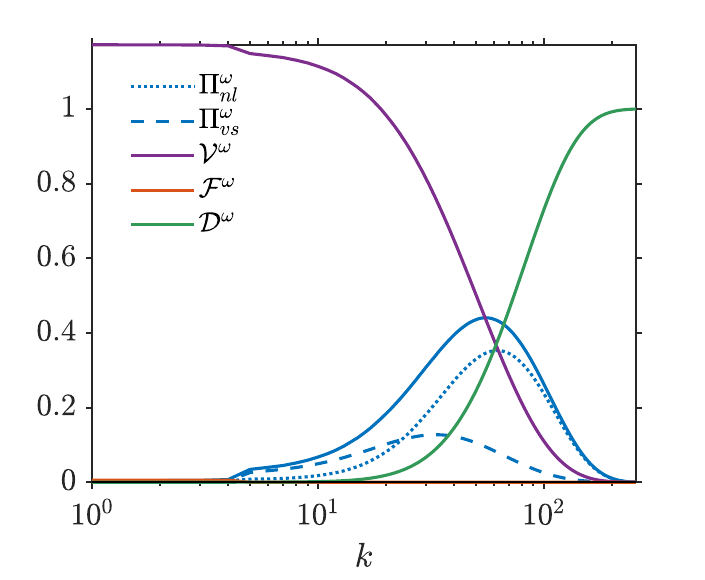}};
\node at ( 0,-5.1){\includegraphics[width=0.35\textwidth]{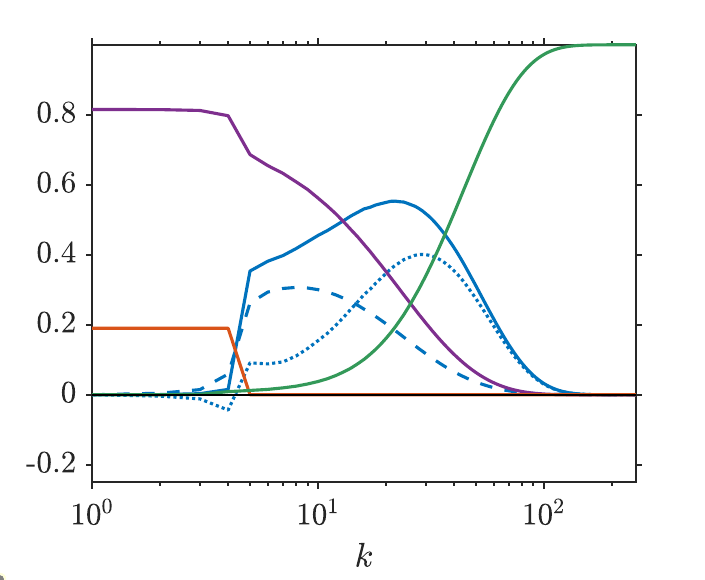}};
\node at ( 6,-5.1){\includegraphics[width=0.35\textwidth]{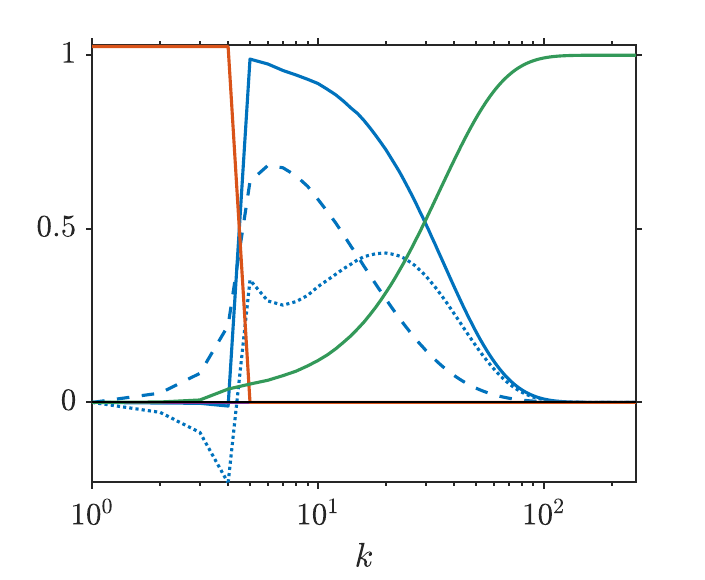}};
\node at (-9,2.5) {(a)};
\node at (-3,2.5) {(b)};
\node at ( 3,2.5) {(c)};
\node at (-9,-2.6){(d)};
\node at (-3,-2.6){(e)};
\node at ( 3,-2.6){(f)};
\end{tikzpicture}
\caption{Evolution of energy and enstrophy budgets under progressive VA suppression. Top panels (a--c) show kinetic energy budgets; bottom panels (d--f) show enstrophy budgets. Columns represent increasing suppression levels: $\gamma = 0$ (a, d), $\gamma = 0.5$ (b, e), and $\gamma = 1$ (c, f). In the enstrophy panels, the solid blue line denotes the total interscale flux, $\Pi^{\omega} = \Pi_{\mathrm{nl}}^{\omega} + \Pi_{\mathrm{vs}}^{\omega}$. As VA is fully suppressed ($\gamma=1$), the classical constant-flux forward energy cascade is completely dismantled (c). Conversely, the total enstrophy flux develops a constant intermediate-scale plateau (f), quantitatively confirming a forward enstrophy cascade. All data refer to the ABC-forced set, with the $\gamma=0$ baseline at $Re_\lambda \approx 120$.}
\label{fig:budget}
\end{figure}

To unravel the physical mechanisms driving the observed spectral modifications and to statistically characterize the transition from a classical energy cascade to an enstrophy cascade, we evaluate the scale-by-scale energy and enstrophy budgets. Fig.~\ref{fig:budget} illustrates these balances and their evolution as VA is progressively suppressed. 
Integrating the modified Lin equation over wavenumbers yields the scale-by-scale energy budget,
\begin{equation}
\mathcal{F}^q(k) + \Pi^q(k) + \gamma \mathcal{V}^q(k) + \mathcal{D}^q(k) = \langle \varepsilon \rangle,
\end{equation}
where the individual scale-dependent contributions are defined as
\begin{align}
\mathcal{F}^q(k) & = \int_k^{\infty} \left[ \oint_{|\bm{k}|=k'} \left( \hat{f}_i \hat{u}_i^* + \hat{f}_i^* \hat{u}_i \right) \mathrm{d}\bm{k} \right] \mathrm{d}k', \\
\Pi^q(k) & = \int_k^{\infty} \left[ \oint_{|\bm{k}|=k'} \left( -\widehat{u_j\partial_j u_i} \hat{u}_i^* - \widehat{u_j\partial_j u_i}^* \hat{u}_i \right) \mathrm{d}\bm{k} \right] \mathrm{d}k', \\
\mathcal{V}^q(k) & = \int_k^{\infty} \left[ \oint_{|\bm{k}|=k'} \left( \hat{g}_i \hat{u}_i^* + \hat{g}_i^* \hat{u}_i \right) \mathrm{d}\bm{k} \right] \mathrm{d}k', \\
\mathcal{D}^q(k) & = \int_0^k \left[ 2 \nu k'^2 E(k') \right] \mathrm{d}k'.
\end{align}
Here, $\hat{\cdot}$ denotes the Fourier transform, the superscript $^*$ indicates complex conjugation, and $\varepsilon = \nu (\partial_j u_i)^2$ is the kinetic energy dissipation rate. A detailed derivation of this scale-by-scale energy budget formulation can be found in \citep{davidson-2004}.

Similarly, the integrated enstrophy budget reads
\begin{equation}
\mathcal{F}^\omega(k) + \Pi_{\mathrm{nl}}^\omega(k) + \Pi_{\mathrm{vs}}^\omega(k) + (1-\gamma)\mathcal{V}^\omega(k) + \mathcal{D}^\omega(k) = \langle \varepsilon_\omega \rangle,
\end{equation}
where the corresponding flux terms are
\begin{align}
\mathcal{F}^\omega(k) & = \int_k^{\infty} \left[ \oint_{|\bm{k}|=k'} \left( \hat{f}^{\omega}_i \hat{\omega}_i^* + \hat{f}^{\omega}_i{}^* \hat{\omega}_i \right) \mathrm{d}\bm{k} \right] \mathrm{d}k', \\
\Pi_{\mathrm{nl}}^\omega(k) & = \int_k^{\infty} \left[ \oint_{|\bm{k}|=k'} \left( -\widehat{u_j\partial_j \omega_i} \hat{\omega}_i^* - \widehat{u_j\partial_j \omega_i}^* \hat{\omega}_i \right) \mathrm{d}\bm{k} \right] \mathrm{d}k', \\
\Pi_{\mathrm{vs}}^\omega(k) & = \int_k^{\infty} \left[ \oint_{|\bm{k}|=k'} \left( \widehat{\text{VT}}_{i} \hat{\omega}_i^* + \widehat{\text{VT}}_{i}^* \hat{\omega}_i \right) \mathrm{d}\bm{k} \right] \mathrm{d}k', \\
\mathcal{V}^\omega(k) & = \int_k^{\infty} \left[ \oint_{|\bm{k}|=k'} \left( \widehat{\text{VA}}_{i} \hat{\omega}_i^* + \widehat{\text{VA}}_{i}^* \hat{\omega}_i \right) \mathrm{d}\bm{k} \right] \mathrm{d}k', \\
\mathcal{D}^\omega(k) & = \int_0^k \left[ 2 \nu k'^2 \Omega(k') \right] \mathrm{d}k'.
\end{align}
In these expressions, $\varepsilon_\omega = \nu (\partial_j \omega_i)^2$ is the enstrophy dissipation rate, and the forcing curl is $f_i^\omega = \epsilon_{i \ell k} \partial_\ell f_k$. The vortex stretching contributions, VA and VT, are explicitly separated.

Physically, the $\mathcal{F}$ terms represent the injection of energy or enstrophy at the large scales via the external forcing, while the $\Pi$ terms quantify the scale-by-scale cascade transfers. In the energy budget, $\Pi^q(k)$ arises entirely from the nonlinear convective term $u_j \partial_j u_i$. In the enstrophy budget, the total interscale transfer, $\Pi^\omega(k) = \Pi_{\mathrm{nl}}^\omega(k) + \Pi_{\mathrm{vs}}^\omega(k)$, is driven by the combined action of advection ($u_j \partial_j \omega_i$) and vortex tilting VT. The $\mathcal{V}$ terms capture the net non-conservative contributions of the nonlinearities to the respective budgets. By construction, these terms vanish for the energy equation at $\gamma=0$ and for the enstrophy equation at $\gamma=1$, consistent with the emergence of their respective inviscid invariants. Finally, the $\mathcal{D}$ terms represent standard viscous dissipation.

In the classical Navier--Stokes limit ($\gamma = 0$), the dynamics are governed by a forward cascade of kinetic energy. Energy is conserved inviscidly, and the nonlinear term purely redistributes this energy from large to small scales, where it is ultimately dissipated ($\Pi^q(0) = \Pi^q(\infty) = 0$). Simultaneously, VA acts as a volumetric source term that generates enstrophy across all active scales, which is then transferred downscale.

As $\gamma$ is increased, the targeted suppression of VA fundamentally reshapes these balances, driving a clear transition from an energy cascade to an enstrophy cascade. In the energy budget, the modified nonlinear contribution acts as a global sink ($\mathcal{V}^q(k) < 0$ at all scales). Consequently, the energy injected at the forcing scale is no longer simply cascaded forward; rather, it is broadly redistributed toward both larger scales ($\Pi^q < 0$ for $k < k_f$) and smaller scales ($\Pi^q > 0$ for $k > k_f$), effectively destroying the classical forward energy cascade.
As VA weakens, the classical enstrophy production mechanism is dismantled. In the limit of complete VA suppression ($\gamma=1$), $\mathcal{V}^\omega = 0$, and enstrophy emerges as the sole relevant inviscid invariant. Under these conditions, the budget provides a definitive statistical characterization of a forward enstrophy cascade. The vortex-tilting contribution, $\Pi_{\mathrm{vs}}^\omega(k)$, becomes overwhelmingly dominant and dictates the interscale enstrophy transfer. Enstrophy is robustly transferred from large to small scales, exhibiting a remarkably constant total flux, $\Pi^\omega(k) \approx \langle \varepsilon_\omega \rangle $, across the intermediate inertial range until it is consumed by viscous dissipation. This statistical plateau in the spectral flux perfectly mirrors the $k^{-3}$ and $k^{-1}$ scalings observed in the intermediate energy and enstrophy spectra.

\subsection{Dissipative Anomalies and Finite-Reynolds Scaling}

\begin{figure}
\centering
\begin{tikzpicture}
\node at (-4.5,0){\includegraphics[width=0.49\textwidth]{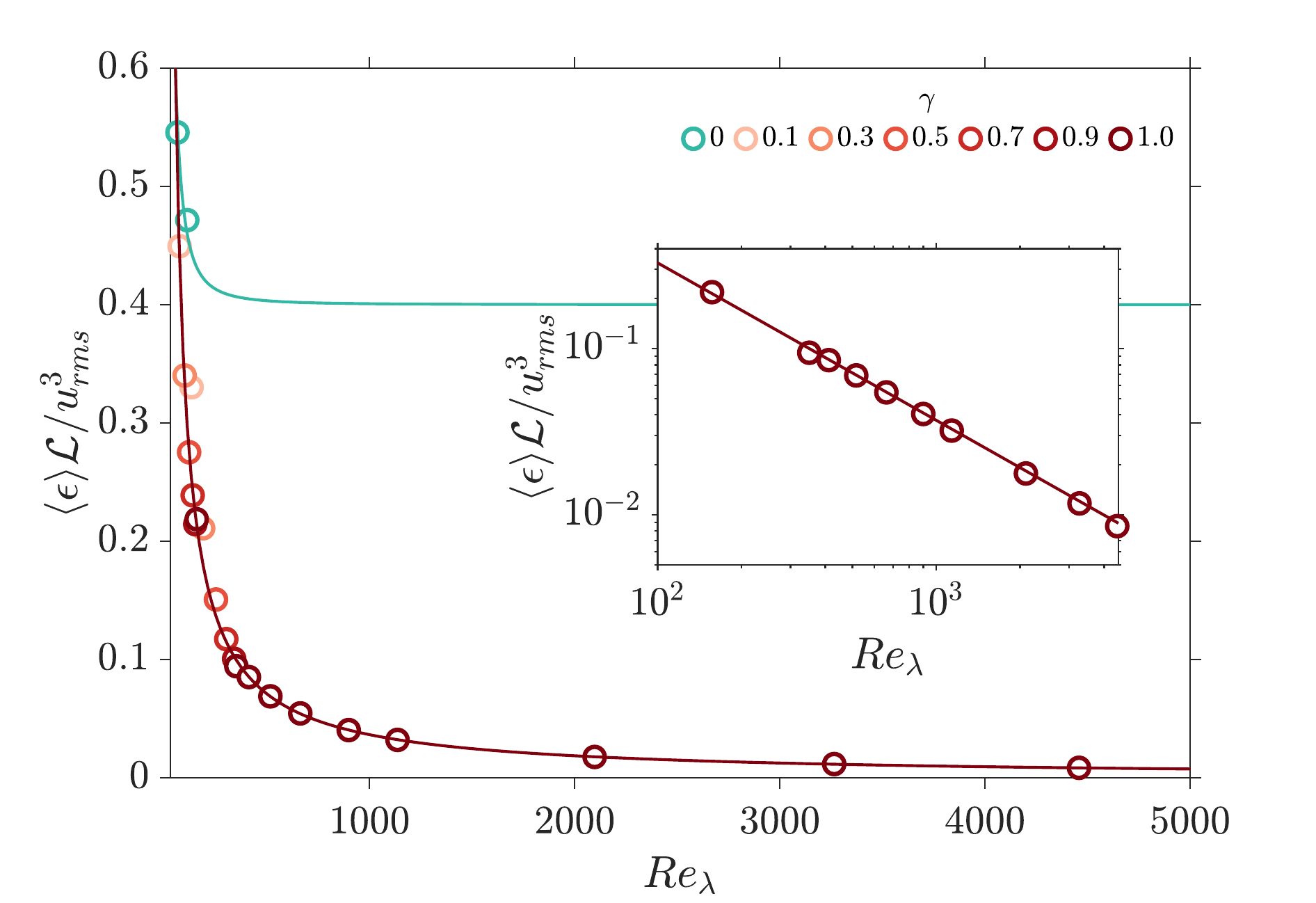}};
\node at ( 4.5,0){\includegraphics[width=0.49\textwidth]{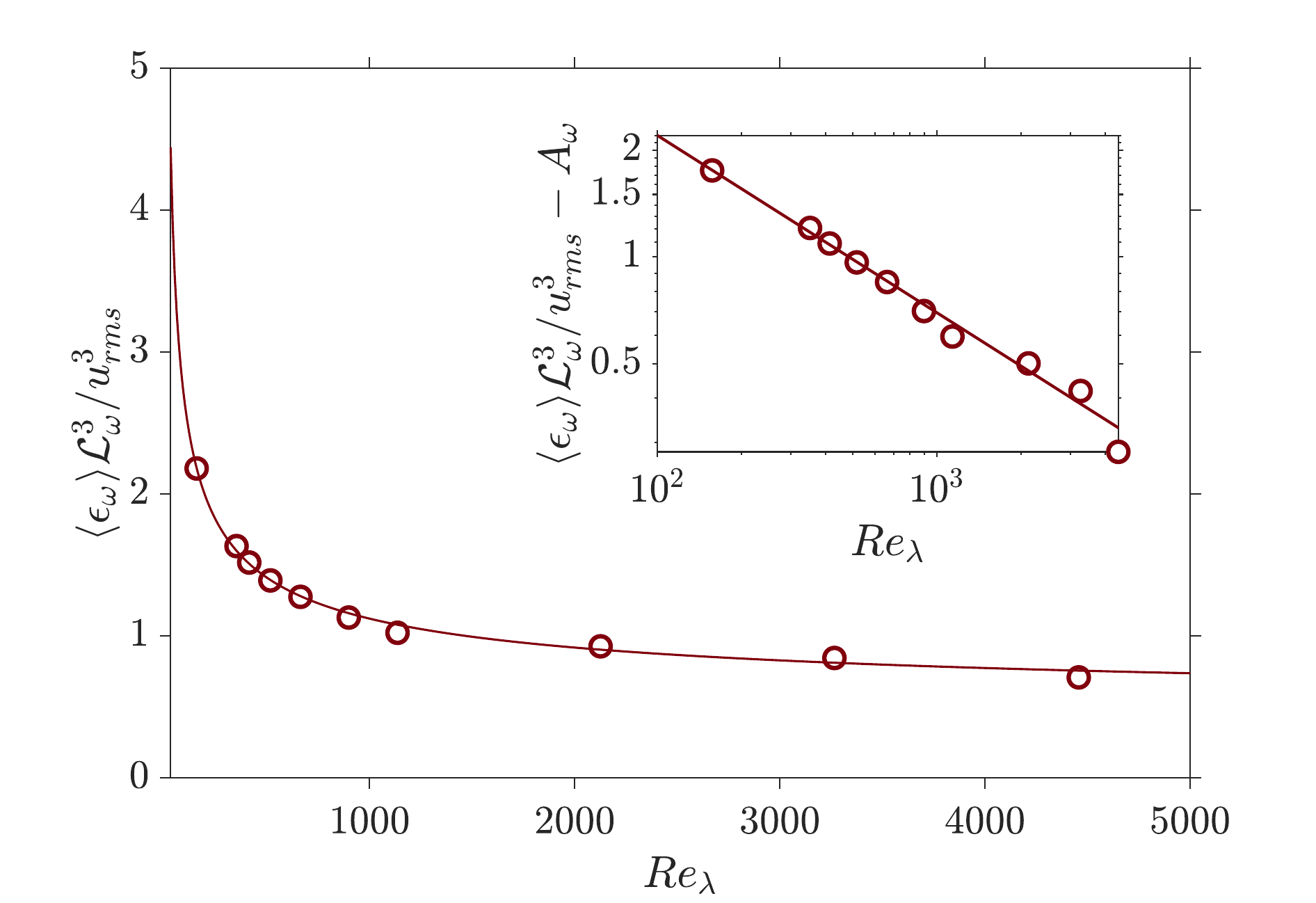}};
\node at (-8.5,2.5){(a)};
\node at ( 0.5,2.5){(b)};
\end{tikzpicture}
\caption{
Evolution of dissipative anomalies under progressive VA suppression. (a) Normalized kinetic-energy dissipation rate, $\beta \equiv \langle \varepsilon \rangle \mathcal{L}/u_{\rm rms}^3$, versus Taylor-scale Reynolds number $Re_\lambda$. Solid green line: phenomenological fit $A[1+\sqrt{1+(BRe_\lambda^{-1})^2}]$ \citep{doering-foias-2002} with $A=0.2, B=92$ \citep{donzis-sreenivasan-yeung-2005}, capturing the classical non-zero anomaly in standard turbulence ($\gamma=0$). Conversely, this anomaly vanishes completely for $\gamma=1$; the solid red line denotes the fit $A + B Re_\lambda^{C}$, yielding an asymptote $A\approx0$, $B \approx 25$, and $C\approx-0.94$, which validates the theoretically predicted $Re_\lambda^{-1}$ decay. (b) Normalized enstrophy dissipation rate, $\beta_\omega \equiv \langle \varepsilon_\omega \rangle \mathcal{L}_\omega^3/u_{\rm rms}^3$, for $\gamma=1$, where $\mathcal{L}_\omega=\frac{\pi}{2\omega_{\rm rms}^2}\int_0^\infty k^{-1}\Omega(k)\,\mathrm{d}k$ defines the enstrophy integral scale. Solid red line: fit $A_\omega + B_\omega Re_\lambda^{C_\omega}$. Convergence to a finite plateau ($A_\omega\approx0.406$, with $B_\omega \approx 22.12$ and $C_\omega\approx-0.49$) confirms the emergence of anomalous enstrophy dissipation, approached via a $Re_\lambda^{-1/2}$ finite-Reynolds-number correction. All data correspond to ABC-forced simulations.
}
\label{fig:dissipation}
\end{figure}

The depletion of small-scale velocity gradients raises a fundamental question: does suppressing VA also suppress the dissipative anomaly, i.e., the persistence of finite energy dissipation in the limit of vanishing viscosity, $\nu \to 0$ \citep{eyink-2024}? 

Fig.~\ref{fig:dissipation}(a) shows the normalized mean kinetic-energy dissipation rate, $\beta \equiv \langle \varepsilon \rangle \mathcal{L}/u_{\rm rms}^3$, as a function of the Taylor-scale Reynolds number $Re_\lambda$, where $\mathcal{L} = \frac{\pi}{2u_{\rm rms}^2} \int_0^\infty k^{-1} E(k)\,\mathrm{d}k$ defines the integral length scale. In standard Navier--Stokes turbulence ($\gamma=0$), the normalized dissipation is known to approach a finite asymptotic value as $Re_\lambda \to \infty$, which is the classical manifestation of the dissipative anomaly \citep{doering-foias-2002,donzis-sreenivasan-yeung-2005}; note, however, that recent evidence suggests a possible slow decay of $\beta$ with $Re$ in flows without solid boundaries \citep{iyer-etal-2025}. Our simulations show that the progressive suppression of VA alters this scaling drastically. As $\gamma$ increases, the dissipation decreases systematically. In the limit of complete suppression ($\gamma=1$), the data strictly follow a $\beta \sim Re_\lambda^{-1}$ decay, marking the absolute disappearance of anomalous energy dissipation. Because enstrophy production is dismantled when VA is removed, the total enstrophy remains bounded, meaning that $\langle \varepsilon \rangle = \nu\langle\omega^2\rangle \to 0$ as $\nu \to 0$.

Nevertheless, while energy dissipation vanishes, an anomalous dissipation emerges for enstrophy, which now acts as the relevant inviscid invariant of the dynamics. Fig.~\ref{fig:dissipation}(b) reports the normalized enstrophy dissipation rate, $\beta_\omega \equiv \langle \varepsilon_\omega \rangle \mathcal{L}_\omega^3/u_{\rm rms}^3$, where $\langle \varepsilon_\omega \rangle = \nu \langle |\bm{\nabla} \bm{\omega}|^2 \rangle = 2\nu P$ (with $P=\int_0^\infty k^4E(k)\,\mathrm{d}k$ being the palinstrophy), and $\mathcal{L}_\omega = \frac{\pi}{2 \omega_{\rm rms}^2} \int_0^\infty k^{-1} \Omega(k)\,\mathrm{d}k$ is the enstrophy integral scale. For $\gamma=1$, the data approach a finite, non-zero asymptotic value as $Re_\lambda \to \infty$. 

This behavior can be explained by phenomenological arguments characteristic of an enstrophy cascade. As the energy spectrum approaches a $k^{-3}$ scaling, the velocity field becomes spatially smooth, implying that the enstrophy transfer is dominated by a non-local mechanism driven by the large-scale strain. Consequently, the relevant advective timescale is no longer the local one, but rather the large-scale turnover time, $\tau \sim \ell/\mathcal{U}$. The global enstrophy dissipation rate can thus be estimated as $\langle \varepsilon_\omega \rangle \sim \langle \omega^2 \rangle / \tau$. Since the vorticity magnitude is governed by the large scales in this regime, we can estimate $\langle \omega^2 \rangle \sim (\mathcal{U}/\ell)^2$. Combining these relations yields $\langle \varepsilon_\omega \rangle \sim \mathcal{U}^3/\ell^3$, which leads directly to the equilibrium scaling law:
\begin{equation}
\beta_\omega = \frac{\langle \varepsilon_\omega \rangle \ell^3}{\mathcal{U}^3} \sim A_\omega,
\label{eq:beta_omega}
\end{equation}
where $A_\omega$ is an $\mathcal{O}(1)$ dimensionless constant.

This anomalous enstrophy dissipation dictates the $Re_\lambda^{-1}$ scaling observed for the energy dissipation. We define the viscous enstrophy scale $\ell_\omega$ through the relation $\langle |\bm{\nabla} \bm{\omega}|^2 \rangle = \ell_\omega^{-2}\langle \omega^2 \rangle$. Recalling that $\langle \varepsilon \rangle = \nu \langle \omega^2 \rangle$, we obtain the exact kinematic relation $\langle \varepsilon_\omega \rangle = \langle \varepsilon \rangle / \ell_\omega^2$. Assuming $\ell_\omega$ scales with the dynamically relevant dissipative scale $\eta_\omega = \nu^{1/2}\langle \varepsilon_\omega \rangle^{-1/6}$ \citep{bos-2021}, we find $\ell_\omega^2 \sim \nu \langle \varepsilon_\omega \rangle^{-1/3}$. Substituting this yields $\langle \varepsilon \rangle \sim \nu \langle \varepsilon_\omega \rangle^{2/3}$. Utilizing the enstrophy dissipative anomaly from Eq.~\eqref{eq:beta_omega} ($\langle \varepsilon_\omega \rangle \sim \mathcal{U}^3/\ell^3$), we obtain $\langle \varepsilon \rangle \sim \nu \mathcal{U}^2/\ell^2$. Consequently, the normalized energy dissipation scales as
\begin{equation}
\beta \equiv \frac{\langle \varepsilon \rangle \ell}{\mathcal{U}^3} \sim \frac{\nu}{\mathcal{U}\ell} = Re^{-1}.
\end{equation}

\begin{figure}
\centering
\begin{tikzpicture}
\node at (-4.5,0){\includegraphics[width=0.49\textwidth]{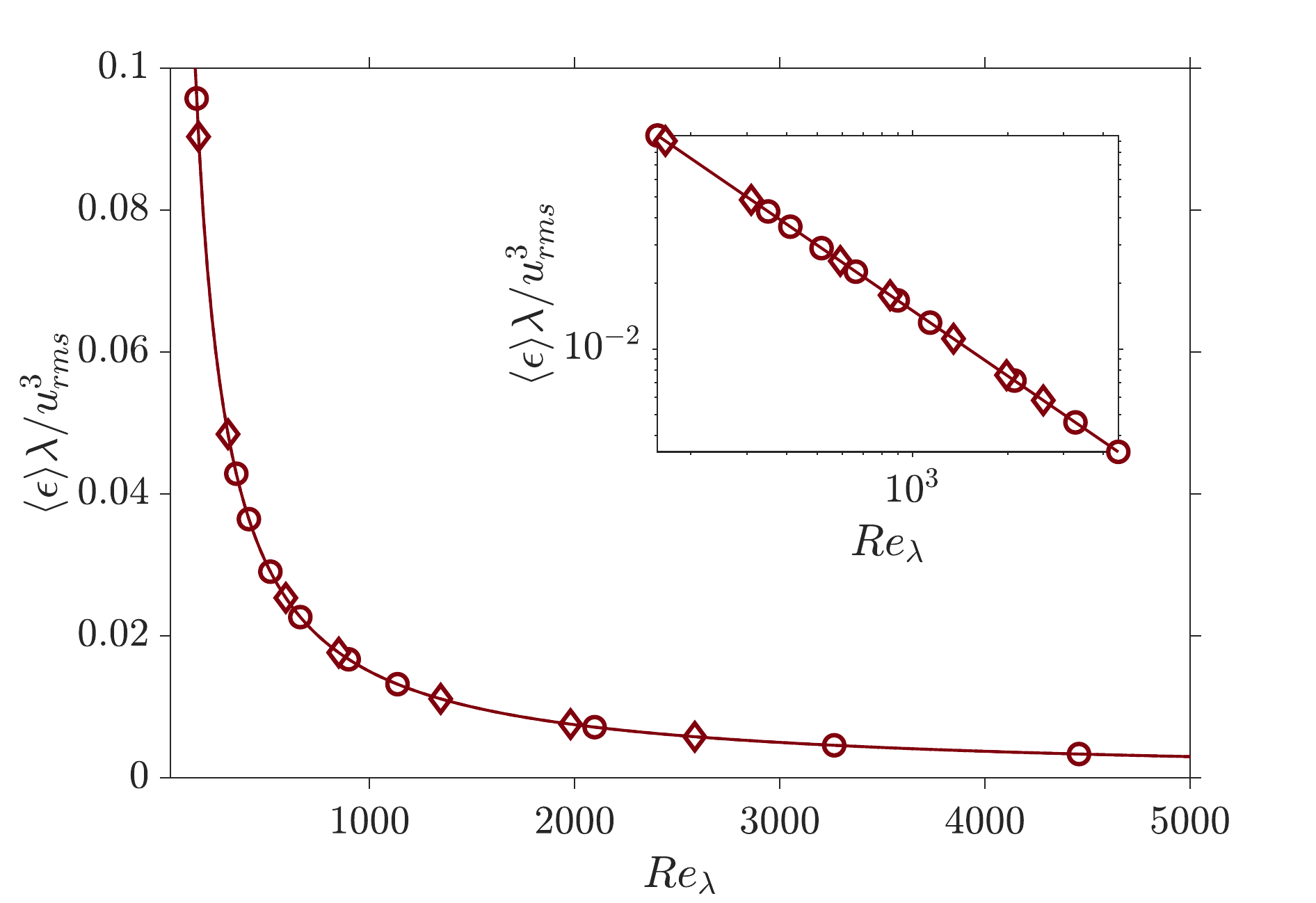}};
\node at ( 4.5,0){\includegraphics[width=0.49\textwidth]{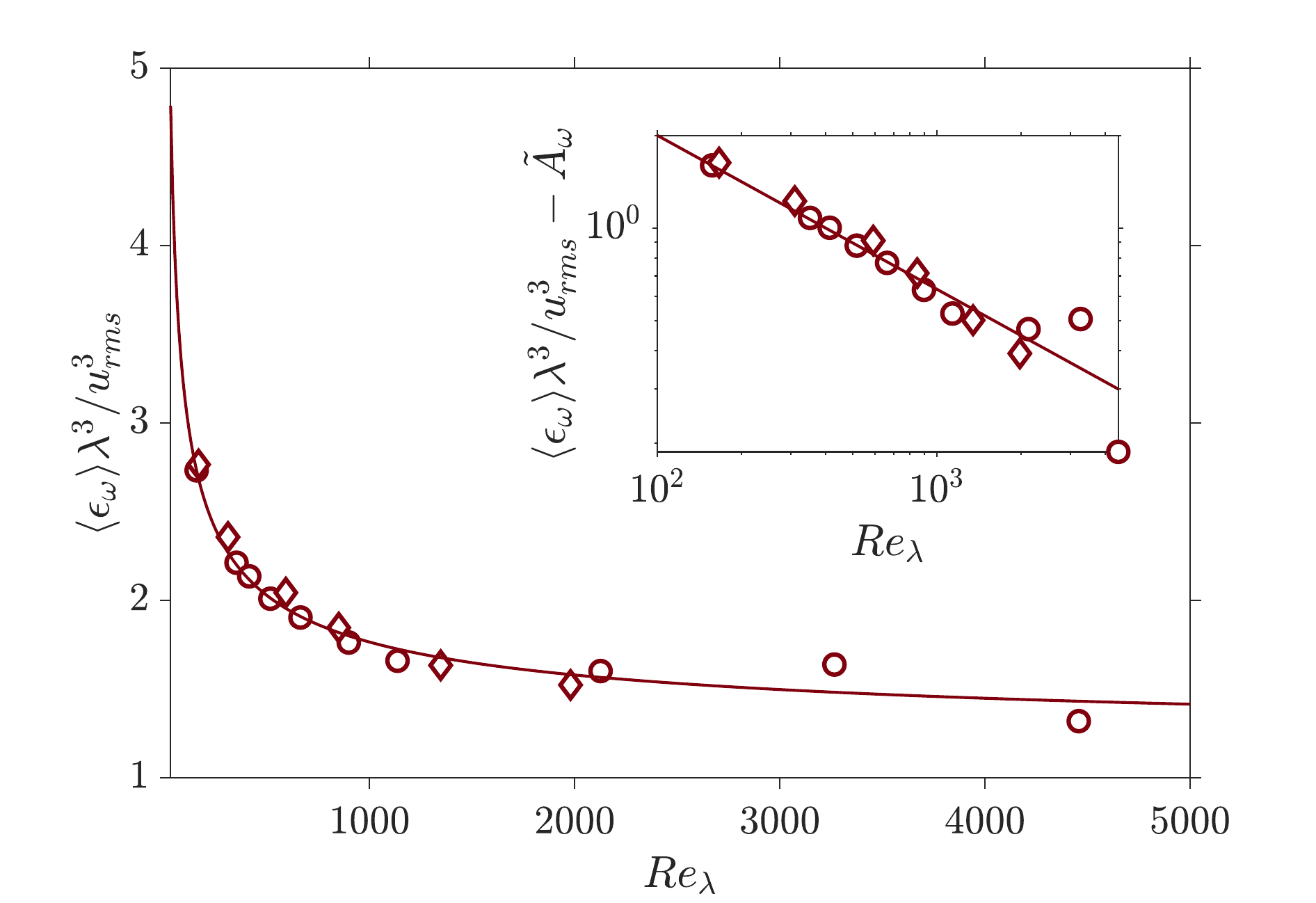}};
\node at (-4.5,-5){\includegraphics[width=0.49\textwidth]{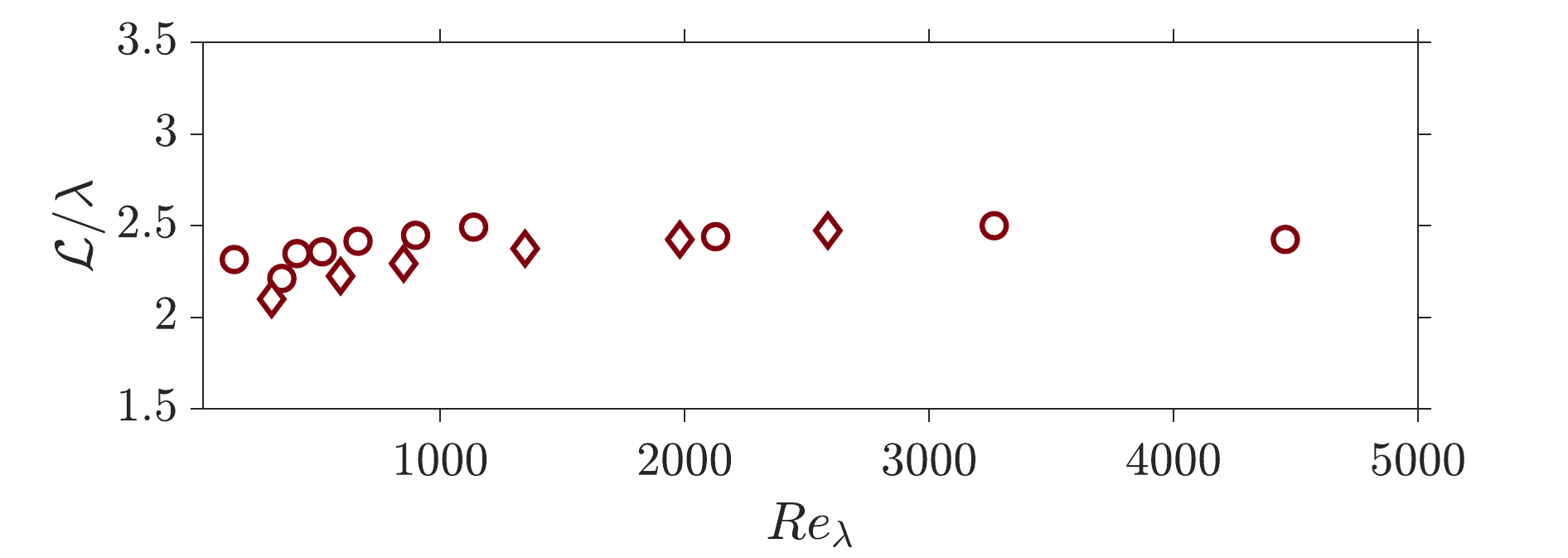}};
\node at ( 4.5,-5){\includegraphics[width=0.49\textwidth]{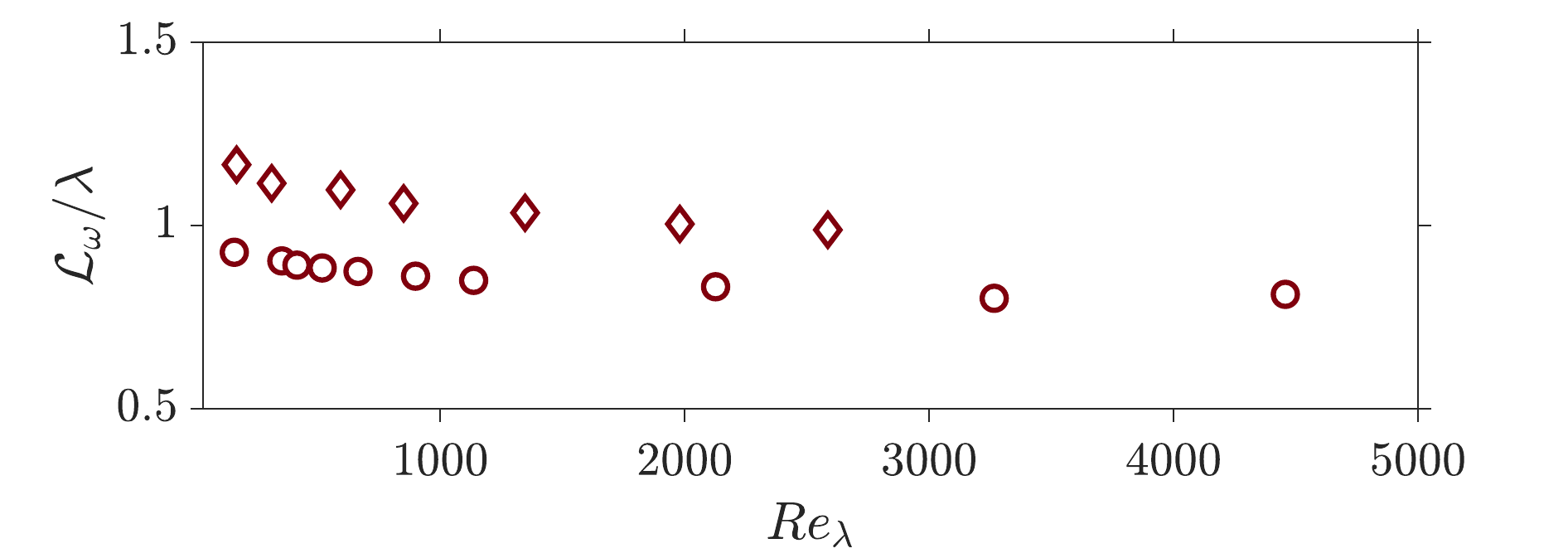}};
\node at (-8.5,2.7){(a)};
\node at ( 0.5,2.7){(b)};
\node at (-8.5,-3.5){(c)};
\node at ( 0.5,-3.5){(d)};
\end{tikzpicture}
\caption{
Universality of dissipative anomalies in the fully VA-suppressed limit ($\gamma=1$). Circles and diamonds denote helical ABC and non-helical (NH) forcing, respectively. To achieve a strict quantitative collapse across datasets, dissipation rates are normalized by the Taylor microscale $\lambda$ rather than integral scales. (a) Normalized kinetic-energy dissipation rate, $\langle \varepsilon \rangle \lambda/u_{\rm rms}^3$, versus $Re_\lambda$. Solid red line: fit $\tilde{A} + \tilde{B} Re_\lambda^{\tilde{C}}$ yielding an asymptote $\tilde{A}\approx0$, $\tilde{B} \approx 15$, and $\tilde{C}\approx-1$, validating the theoretically predicted $Re_\lambda^{-1}$ decay regardless of forcing scheme. (b) Normalized enstrophy dissipation rate, $\langle \varepsilon_\omega \rangle \lambda^3/u_{\rm rms}^3$. Solid red line: fit $\tilde{A}_\omega + \tilde{B}_\omega Re_\lambda^{\tilde{C}_\omega}$. Convergence to a universal finite plateau ($\tilde{A}_\omega\approx 1.1$, with $\tilde{B}_\omega \approx 16.7$ and $\tilde{C}_\omega\approx-0.46$) confirms anomalous enstrophy dissipation approached via a $Re_\lambda^{-1/2}$ finite-Reynolds-number correction. Length-scale ratios (c) $\mathcal{L}/\lambda$ and (d) $\mathcal{L}_\omega/\lambda$. High-$Re_\lambda$ plateaus confirm the kinematic relation $\lambda \sim \mathcal{L} \sim \mathcal{L}_\omega$; however, exact asymptotic proportionality coefficients depend on the forcing's spatial structure, necessitating the $\lambda$-normalization in (a, b).
}
\label{fig:diss-forces}
\end{figure}
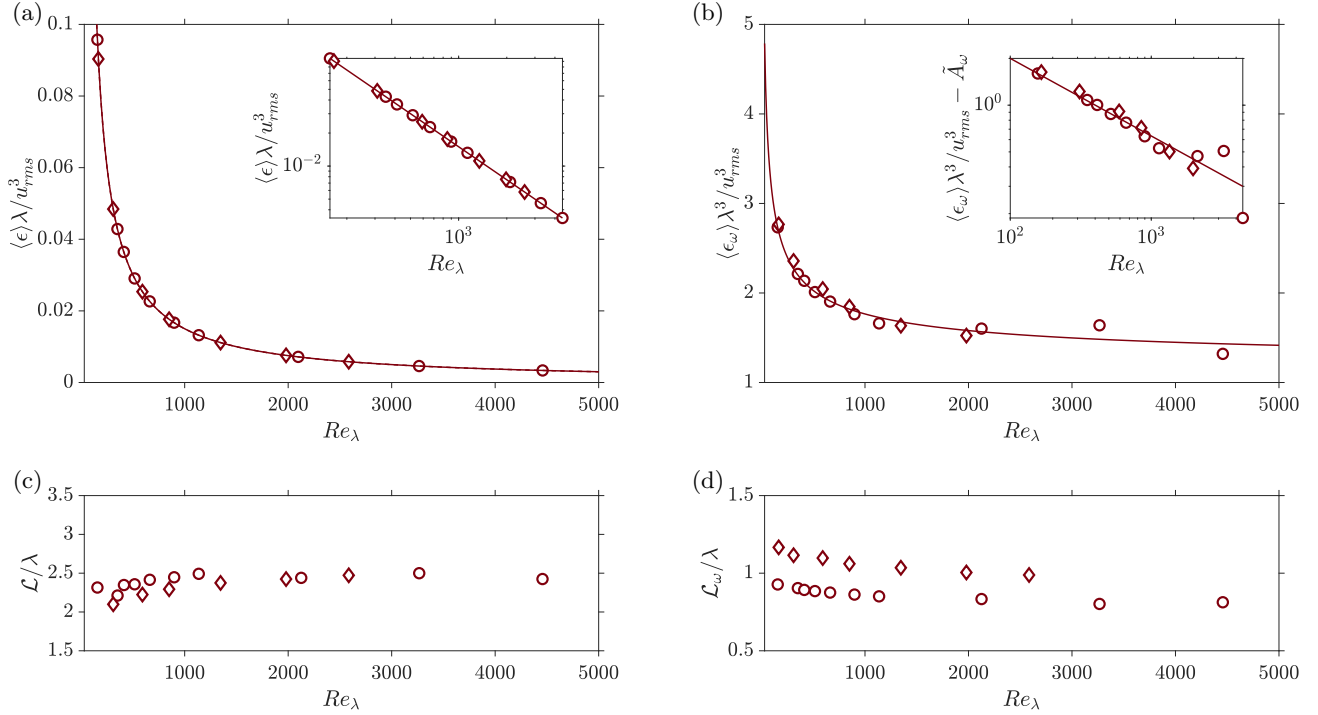

To rigorously link this $Re^{-1}$ scaling to the observed $Re_\lambda^{-1}$ decay in Fig.~\ref{fig:dissipation}(a), we must prove that $Re \sim Re_\lambda$ in our modified system. In classical turbulence, the Taylor microscale $\lambda$ is an intermediate inertial-range scale, leading to $Re_\lambda \sim Re^{1/2}$. However, in the absence of VA, energy and enstrophy remain tightly concentrated at the large scales, heavily suppressing the high-wavenumber tails (as previously shown in Fig.~\ref{fig:spectra}). By estimating the spectral integrals around the dominant large-scale wavenumber $k_c \sim \ell^{-1}$, such that $\int E(k)\,\mathrm{d}k \sim E_c$, $\int k^2 E(k)\,\mathrm{d}k \sim k_c^2 E_c$ and so forth, we find:
\begin{equation}
\lambda^2 = \frac{15\nu \mathcal{U}^2}{\langle \varepsilon \rangle} \sim \frac{\int E(k)\,\mathrm{d}k}{\int k^2E(k)\,\mathrm{d}k} \sim k_c^{-2}.
\end{equation}
Applying the same logic to the integral scales yields $\mathcal{L} \sim k_c^{-1}$ and $\mathcal{L}_\omega \sim k_c^{-1}$. Consequently, the characteristic length scales collapse such that $\lambda \sim \mathcal{L} \sim \mathcal{L}_\omega \sim \ell$ (see Fig.~\ref{fig:diss-forces}(c,d)). Alternatively, substituting the viscous estimate $\langle \varepsilon \rangle \sim \nu \mathcal{U}^2/\ell^2$ into the formal definition of the Taylor microscale directly yields $\lambda^2 \sim \ell^2$. Because the Taylor microscale remains dynamically tied to the large energy-containing scales, the Reynolds numbers are strictly proportional: $Re_\lambda = \mathcal{U}\lambda/\nu \sim \mathcal{U}\ell/\nu = Re$. This constraint establishes that $\beta \sim Re_\lambda^{-1}$, in perfect agreement with the numerical data.

Finally, we address the finite-$Re_\lambda$ corrections observed in the enstrophy dissipation data. Our numerical results suggest a scaling of the form $\beta_\omega = A_\omega + B_\omega Re_\lambda^{-1/2}$. Phenomenologically, we hypothesize that these corrections are controlled by the scale separation parameter $\chi \equiv \ell_\omega/\ell$. Using $\ell_\omega \sim \nu^{1/2} (\beta_\omega \mathcal{U}^3/\ell^3)^{-1/6}$, we find:
\begin{equation}
\chi \sim \left( \frac{\nu}{\mathcal{U}\ell} \right)^{1/2} \beta_\omega^{-1/6} = Re_\lambda^{-1/2}\beta_\omega^{-1/6}.
\end{equation}
At leading order, $\beta_\omega \to A_\omega$, meaning $\chi$ vanishes proportionally to $Re_\lambda^{-1/2}$. By performing a regular Taylor expansion of $\beta_\omega$ in powers of $\chi$ ($\beta_\omega = a_0 + a_1\chi + a_2\chi^2 + \dots$), we recover the scaling $\beta_\omega = b_0 + b_1 Re_\lambda^{-1/2} + b_2 Re_\lambda^{-1} + \dots$, thereby explaining the dominant $Re_\lambda^{-1/2}$ correction.

To establish the universality of these asymptotic limits, we compare the dissipation scalings produced by the ABC forcing (circles) against those from the non-helical (NH) forcing scheme (diamonds) in Fig.~\ref{fig:diss-forces}. Crucially, both the robust $Re_\lambda^{-1}$ decay of the energy dissipation and the emergence of an anomalous enstrophy dissipation, alongside its associated $Re_\lambda^{-1/2}$ finite-Reynolds correction, are consistently recovered under the NH forcing. 
However, a strict quantitative collapse of the two datasets, i.e., the exact matching of the asymptotic constants $A_\omega$ and $B_\omega$, is achieved only when the dissipation rates are normalized using the Taylor microscale, $\lambda$, rather than the integral scales, $\mathcal{L}$ or $\mathcal{L}_\omega$. As demonstrated previously, the relation $\lambda \sim \mathcal{L} \sim \mathcal{L}_\omega$ holds true for both configurations. Nevertheless, the exact proportionality coefficients connecting these scales are non-universal; they depend on the spatial structure and helicity of the large-scale energy injection, as evidenced by the differing plateau values of $\mathcal{L}/\lambda$ and $\mathcal{L}_\omega/\lambda$ in Fig.~\ref{fig:diss-forces}(c,d). Normalizing by $\lambda$ effectively absorbs these forcing-dependent prefactors. This resulting collapse unequivocally confirms that the anomalous dissipation limits observed here are intrinsic, universal properties of the VA-suppressed Navier--Stokes dynamics, entirely independent of the specific large-scale forcing mechanism.

\subsection{Anomalous Scaling and Multifractal Intermittency}
\label{sec:intermittency}

\begin{figure}
\centering
\begin{tikzpicture}
\node at (-4.5,0){\includegraphics[width=0.49\textwidth]{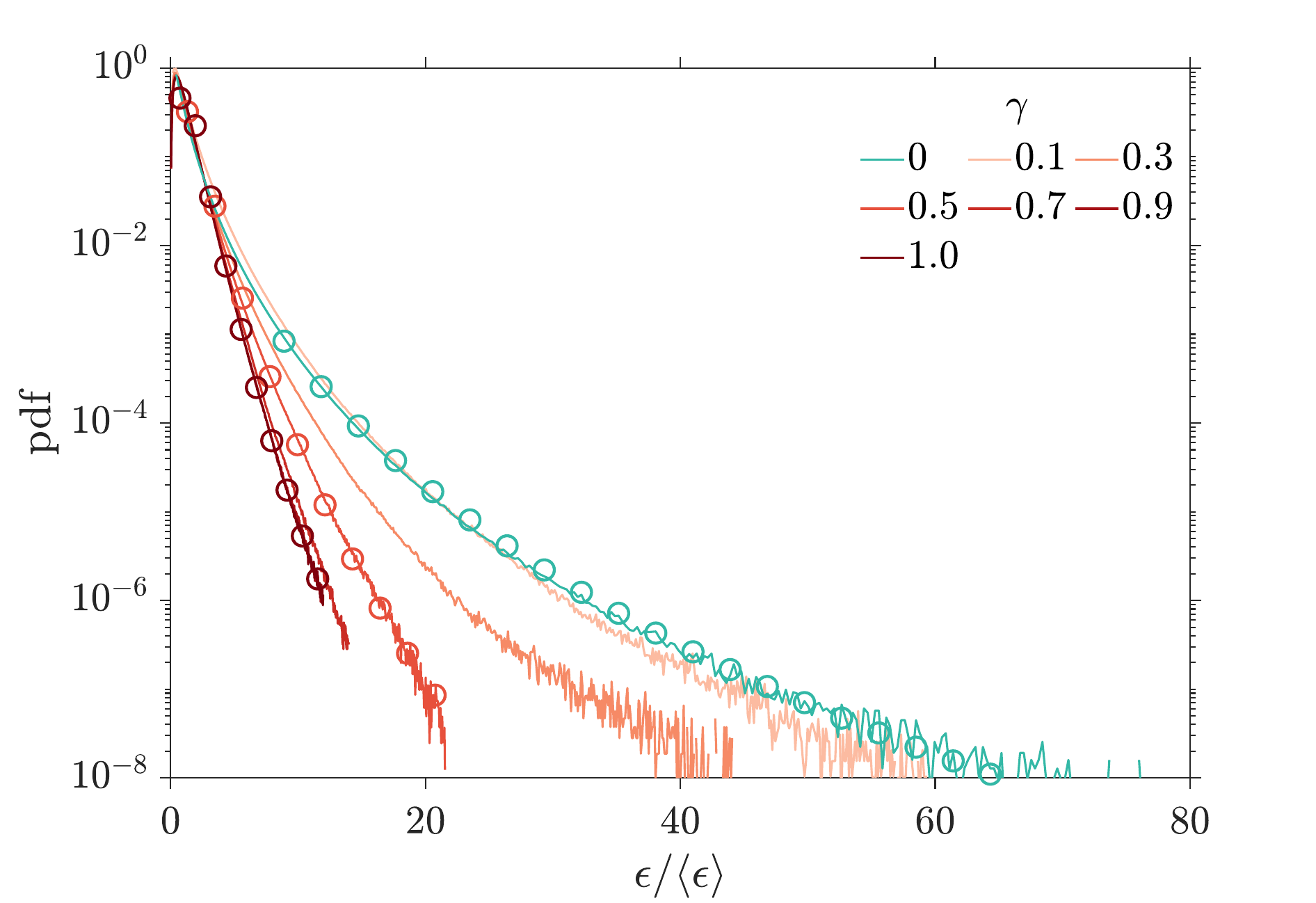}};
\node at ( 4.5,0){\includegraphics[width=0.49\textwidth]{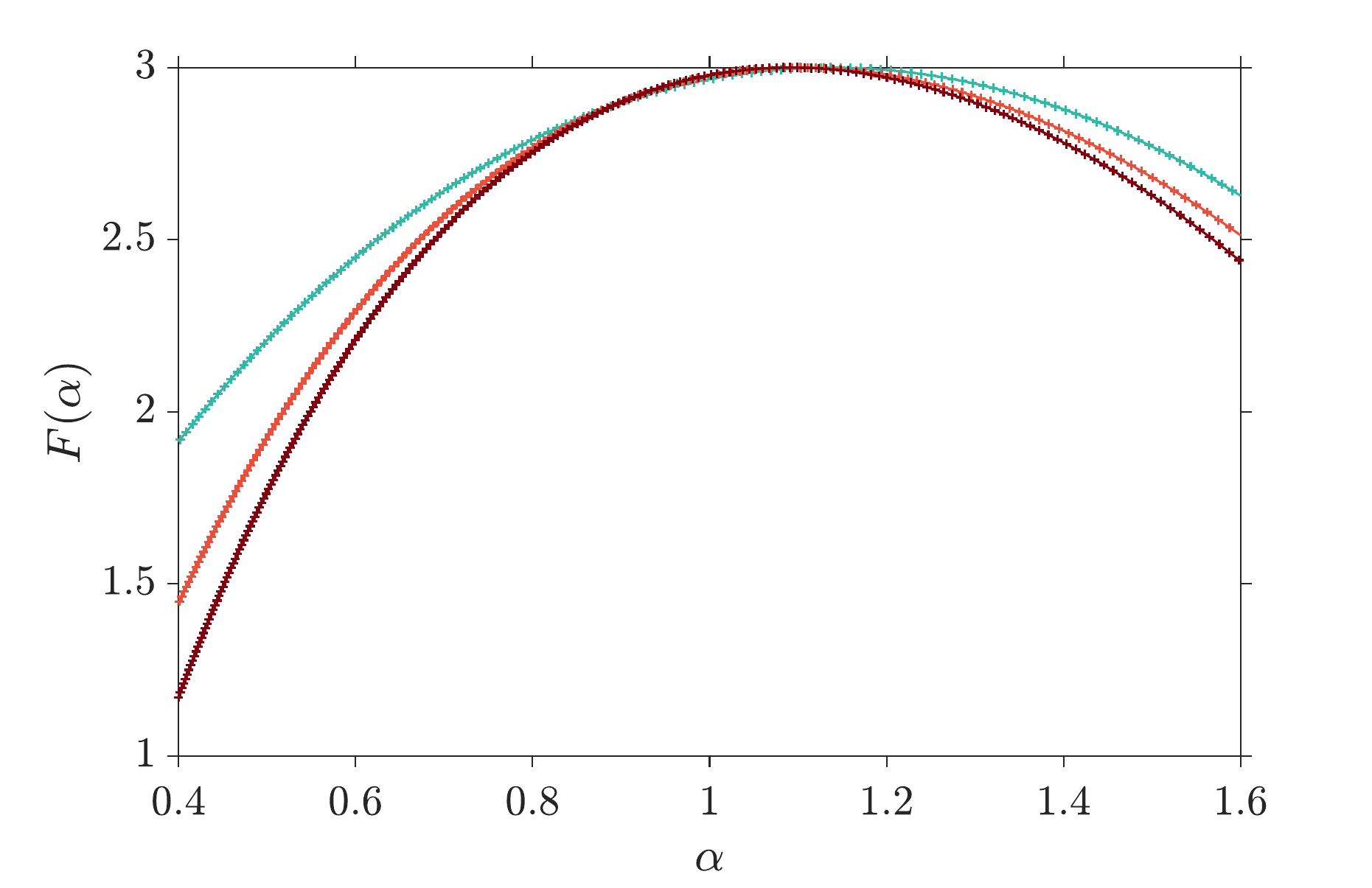}};
\node at (-8.5,2.5){(a)};
\node at ( 0.5,2.5){(b)};
\end{tikzpicture}
\caption{
Impact of VA suppression on intermittency via dissipation statistics. (a) Probability density functions (PDFs) of the normalized energy dissipation rate for varying $\gamma$. Circles: log-normal fits, $p(\varepsilon) = c_1 \exp[-c_2(\log \varepsilon)^2]$. Increasing $\gamma$ severely depletes the heavy tails, indicating a stark reduction in extreme fluctuation events. Fit parameters $(c_1, c_2)$ are $(0.06, 0.89)$ for standard dynamics ($\gamma=0$), $(0.36, 1.66)$ for partial ($\gamma=0.5$), and $(0.56, 2.11)$ for complete VA suppression ($\gamma=1$), quantitatively confirming the narrowed distribution while demonstrating a surviving log-normal core. (b) Multifractal spectrum $F(\alpha)$ of the coarse-grained energy dissipation for $\gamma=0$, $0.5$ and $1$. Although the spectrum narrows upon VA removal, a well-defined singularity range persists at $\gamma=1$, confirming that VT alone sustains multifractal intermittency. Data correspond to ABC-forced simulations at a baseline $Re_\lambda \approx 120$.
}
\label{fig:diss-intermittency}
\end{figure}
The suppression of VA profoundly alters the intermittent nature of the flow, a phenomenon we first illustrate through the statistics of the energy dissipation (Fig.~\ref{fig:diss-intermittency}). As VA is progressively reduced, the probability density function (PDF) of the dissipation becomes increasingly narrow (Fig.~\ref{fig:diss-intermittency}a). The heavy tails characteristic of classical turbulence are severely suppressed, indicating a smoother small-scale velocity field and a stark depletion of extreme fluctuation events. Nevertheless, the core of the dissipation PDF remains approximately log-normal across all values of $\gamma$, including the limit of complete VA suppression ($\gamma=1$). This provides the first indication that VT alone is sufficient to sustain intermittent fluctuations, albeit with substantially reduced intensity.

This residual intermittency is further corroborated by analyzing the statistics of the coarse-grained energy dissipation \citep{meneveau-sreenivasan-1991,frisch-1996}.
We compute the $q$-th moment of the locally averaged dissipation field, $\varepsilon_r \equiv 2\nu \langle s_{ij}s_{ij}\rangle_r$, where $\langle\cdot\rangle_r$ denotes the spatial average over a sphere of radius $r$.
In the inertial range, these moments exhibit the scaling behavior $\langle \varepsilon_r^q \rangle \sim r^{\lambda_q}$. The Legendre transform of these scaling exponents defines the multifractal spectrum $F(\alpha)$:
\begin{equation}
\lambda_q = \inf_{\alpha} \left[ q(\alpha - 1) + 3 - F(\alpha) \right],
\end{equation}
where singularities of strength $\alpha-1$ occupy sets of fractal dimension $F(\alpha)$. 
Fig.~\ref{fig:diss-intermittency}(b) displays $F(\alpha)$ for $\gamma=0$, $0.5$, and $1$.
In standard Navier--Stokes turbulence ($\gamma=0$), the spectrum is broad and consistent with previous literature \citep{meneveau-sreenivasan-1991}. Upon complete VA suppression ($\gamma=1$), the spectrum narrows, mathematically reflecting the reduced intermittency observed in the PDFs. Nevertheless, the flow retains a well-defined multifractal spectrum spanning a distinct range of singularities. This demonstrates that, despite the smoothed small-scale dynamics, VT alone is mechanically sufficient to sustain multifractal intermittency.

\begin{figure}
\centering
\begin{tikzpicture}
\node at (-4.5,0){\includegraphics[width=0.49\textwidth]{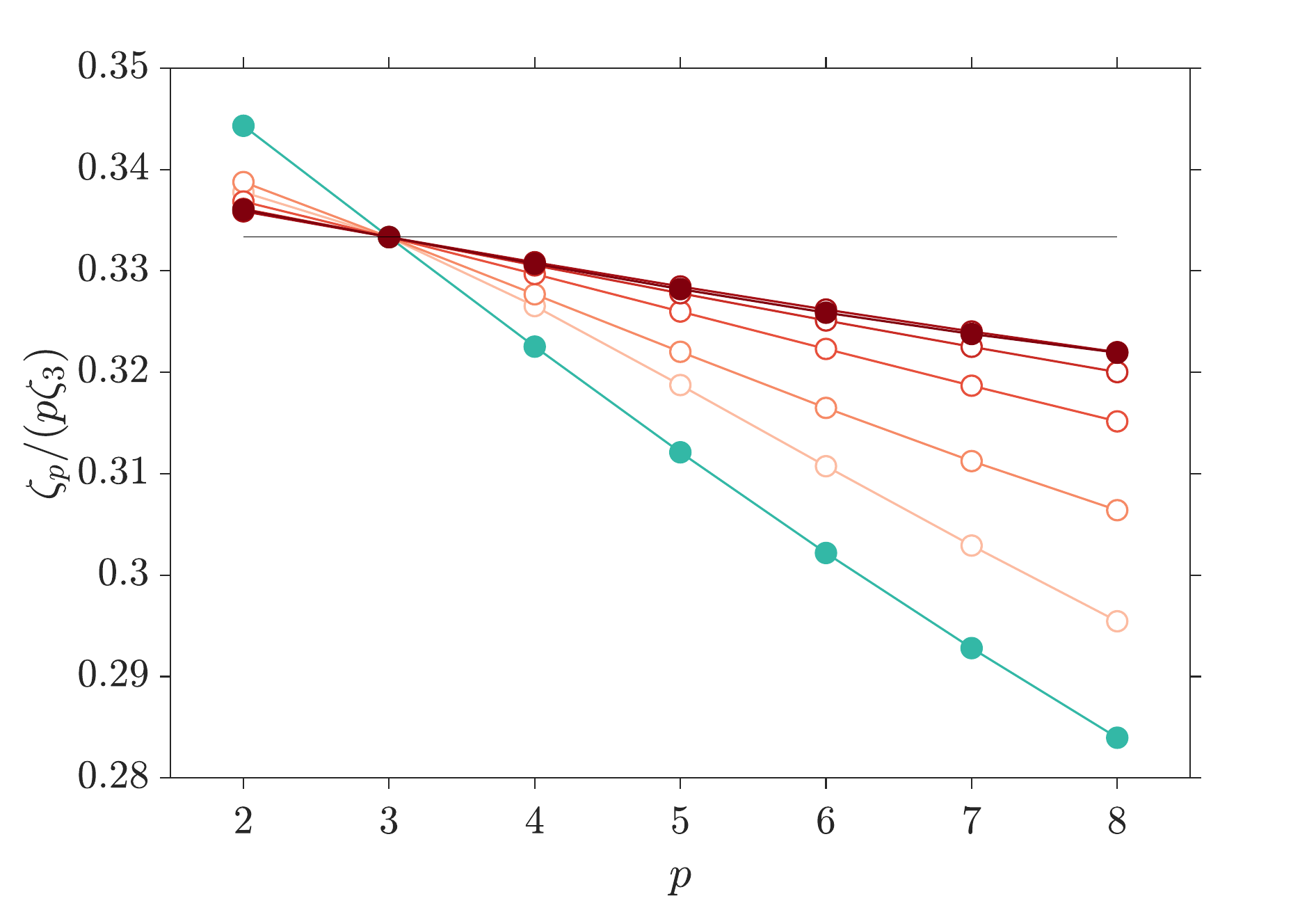}};
\node at ( 4.5,0){\includegraphics[width=0.49\textwidth]{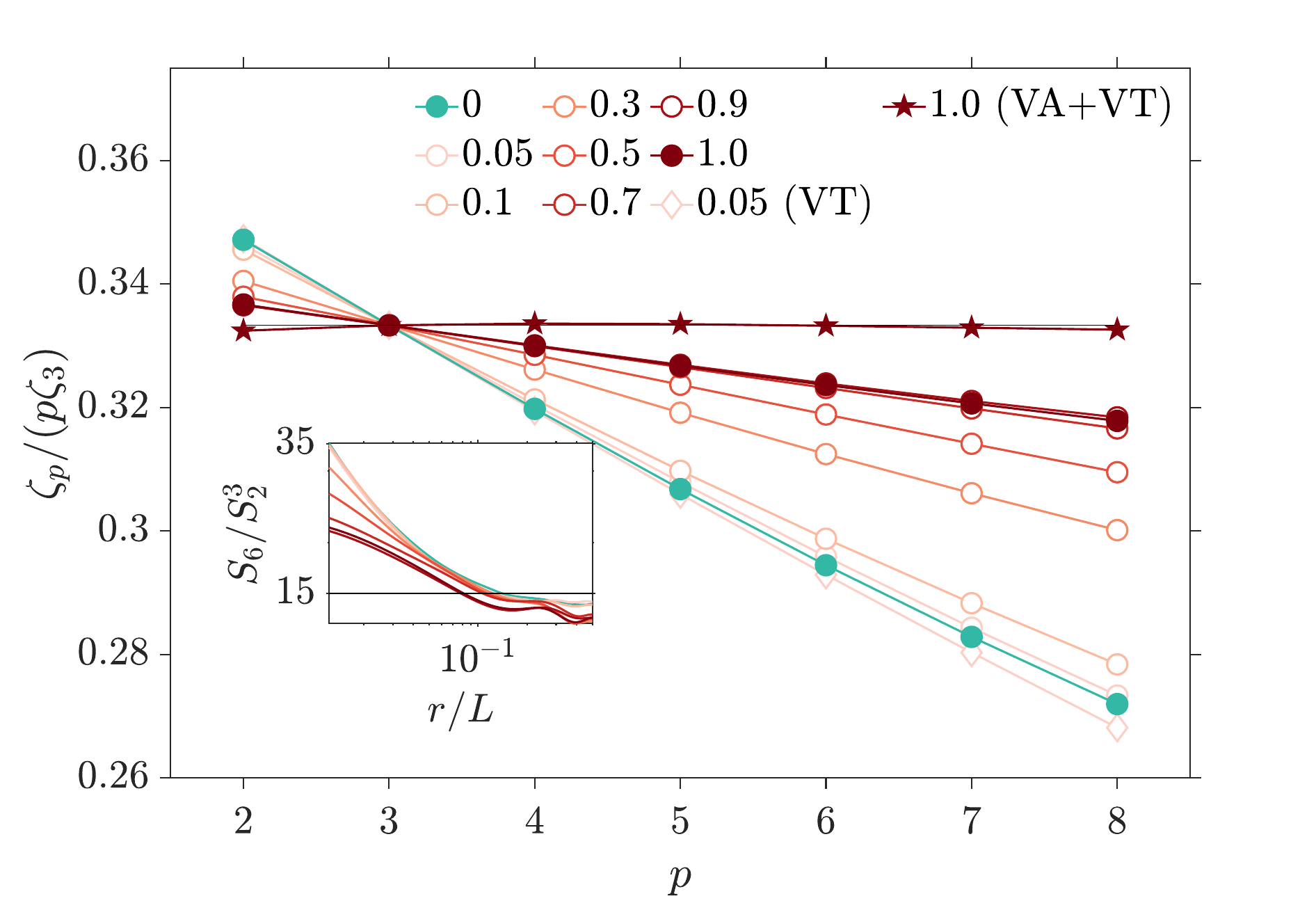}};
\node at (-4.5,-6){\includegraphics[width=0.49\textwidth]{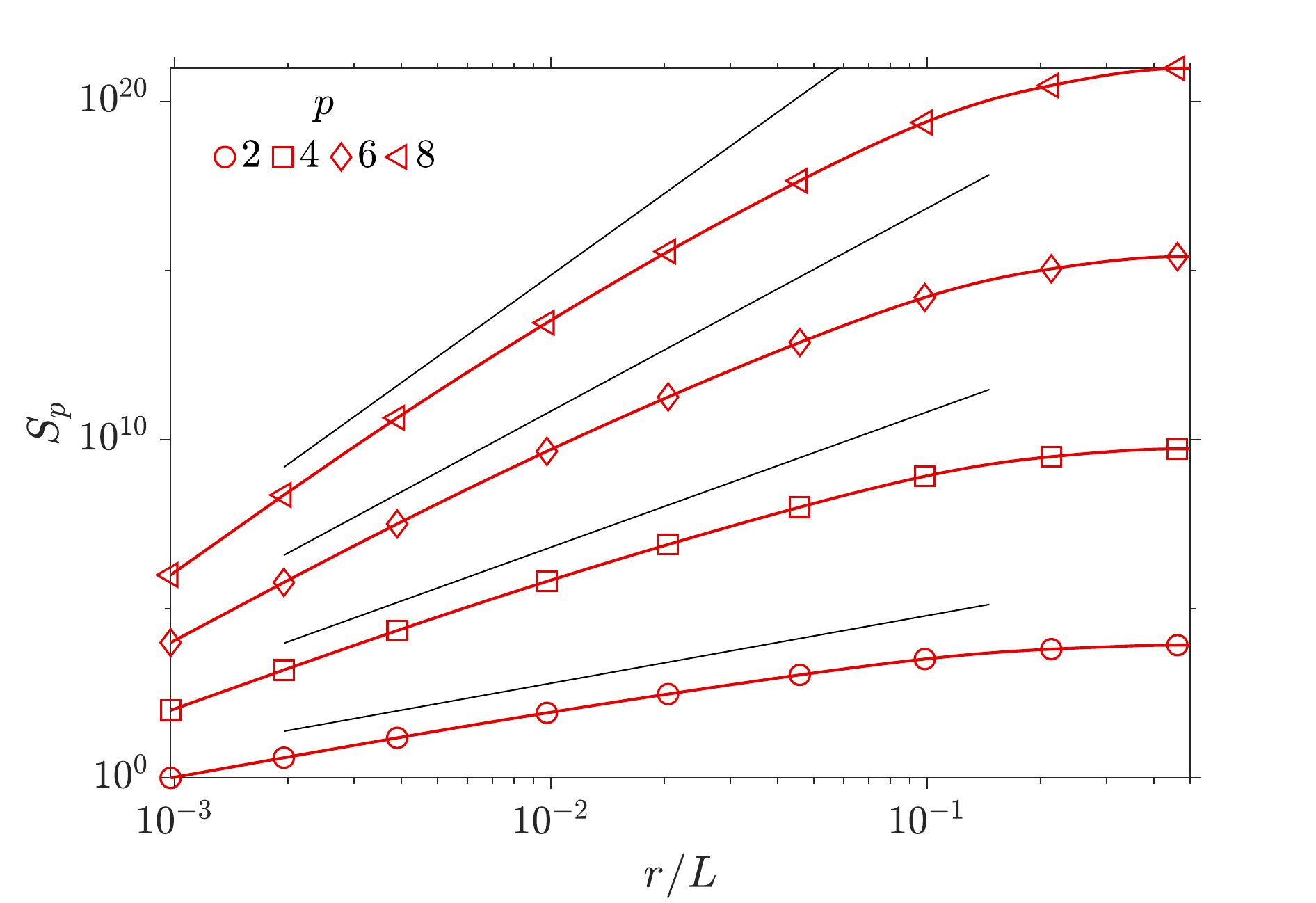}};
\node at ( 4.5,-6){\includegraphics[width=0.49\textwidth]{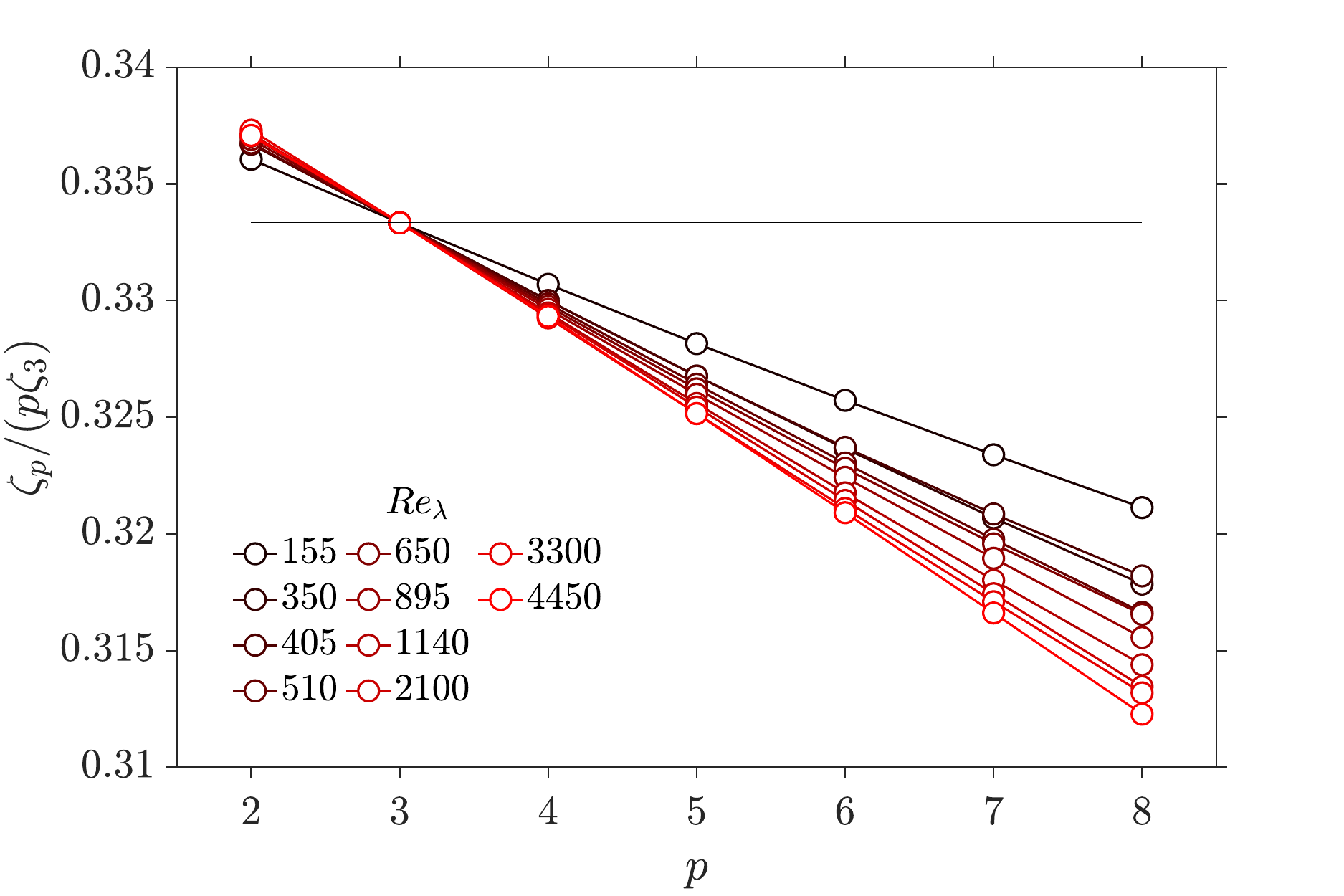}};
\node at (-8.5,2.9){(a)};
\node at ( 0.5,2.9){(b)};
\node at (-8.5,-3.1){(c)};
\node at ( 0.5,-3.1){(d)};
\end{tikzpicture}
\caption{
Kinematic signatures of intermittency and anomalous scaling in the velocity structure functions under VA suppression. (a, b, d) Effective scaling ratio $\zeta_p / (p \zeta_3)$ for longitudinal structure functions $S_p(r)$, extracted via extended self-similarity \citep{benzi-etal-1993}. (a, b) Evolution of scaling ratios for varying $\gamma$ under ABC forcing at baseline $Re_\lambda \approx 65$ (a) and $Re_\lambda \approx 120$ (b). Panel (b) isolates mechanistic origins: circles denote progressive VA suppression; red stars mark complete vortex-stretching suppression (VA+VT, $\gamma=1$), which perfectly restores the non-intermittent dimensional scaling ($1/3$); orange diamonds indicate targeted VT suppression ($\gamma=0.05$), exacerbating anomalous deviations. Inset: Flatness-like ratio $S_6/S_2^3(r)$ versus $r$. Deviations from the Gaussian value of 15 strictly persist across all $\gamma$, confirming residual intermittency. (c) Raw $S_p(r)$ for $p=2, 4, 6, 8$ in the fully VA-suppressed limit ($\gamma=1$) at $Re_\lambda \approx 3300$. Solid lines denote the $r^p$ dimensional scaling predicted by the enstrophy cascade. (d) Reynolds-number dependence of $\zeta_p / (p \zeta_3)$ for the completely VA-suppressed state ($\gamma=1$) under ABC forcing, confirming robust residual intermittency at high $Re_\lambda$.
}
\label{fig:intermittency}
\end{figure}
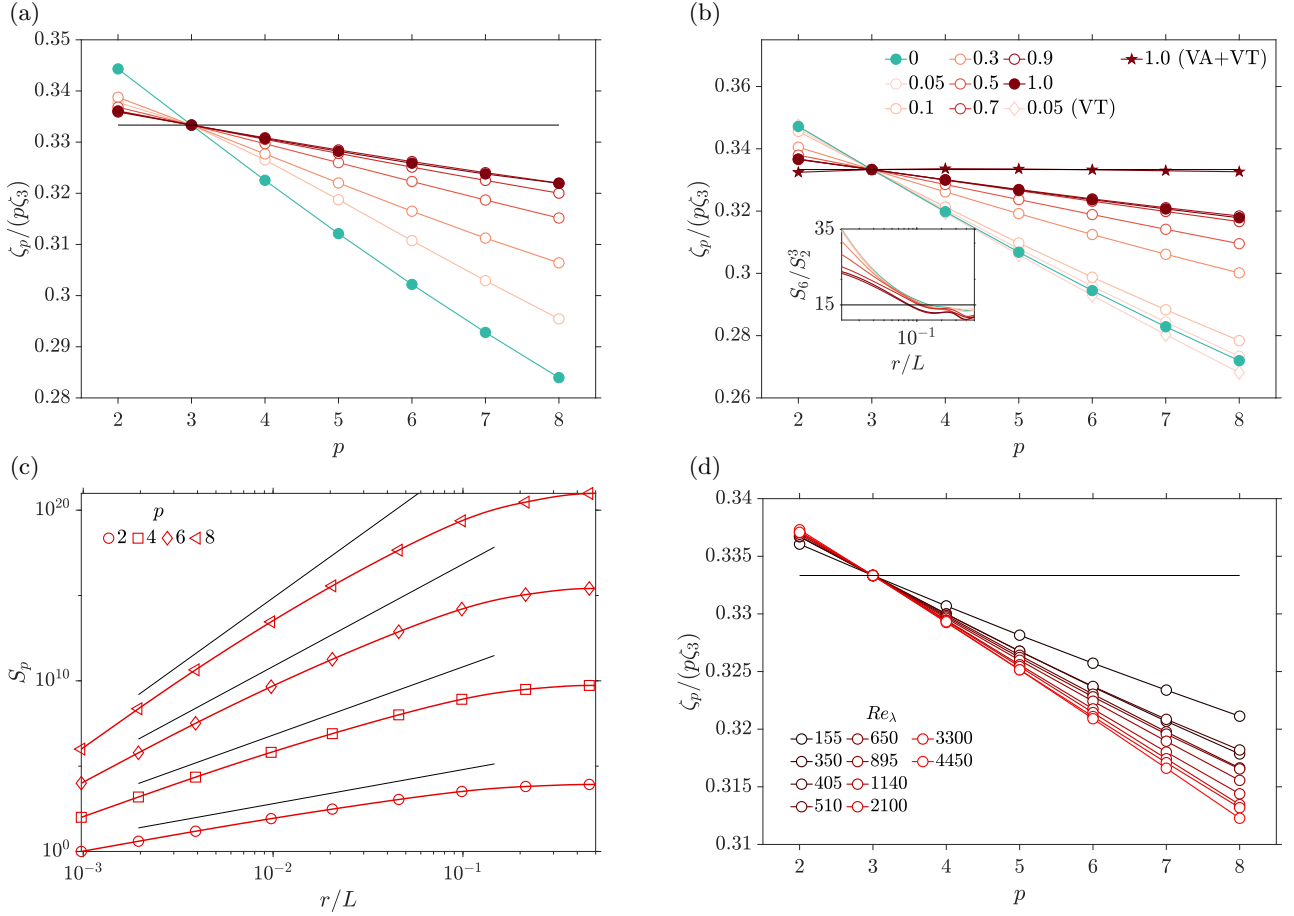

To quantify this intermittency kinematically, we turn to the higher-order longitudinal velocity structure functions, $S_p(r) \equiv \langle [\delta u_\parallel(r)]^p \rangle \sim r^{\zeta_p}$, where $\delta u_\parallel$ is the longitudinal velocity increment. In classical three-dimensional turbulence ($\gamma=0$), the scaling exponents $\zeta_p$ exhibit a nonlinear dependence on $p$, deviating clearly from the inertial dimensional prediction of $p/3$. Conversely, in our modified $\gamma=1$ system, dimensional analysis based on the enstrophy cascade predicts a leading-order kinematic scaling $S_p(r) \sim \langle \varepsilon_\omega \rangle^{p/3} r^p$ (shown as reference lines in Fig.~\ref{fig:intermittency}c). The critical question is whether anomalous deviations from this dimensional scaling persist in the complete absence of VA.

Although the direct extraction of scaling exponents from $S_p(r)$ is hindered by logarithmic corrections intrinsic to $k^{-3}$ spectra in the limit of complete VA suppression, we circumvent these kinematic artifacts by computing relative scaling exponents. Figs.~\ref{fig:intermittency}a and \ref{fig:intermittency}b presents the effective scaling ratio $\zeta_p/(p\zeta_3)$, extracted using extended self-similarity (ESS) with respect to $S_3(r)$ \citep{benzi-etal-1993}, for different values of $\gamma$. In the $\gamma=0$ baseline, the exponents show classic anomalous scaling: $\zeta_p/(p\zeta_3) < 1/3$ for $p>3$ and $\zeta_p/(p\zeta_3) > 1/3$ for $p<3$ \citep{frisch-1996}. As $\gamma$ increases, these deviations systematically decrease, reinforcing the picture of a smoothing flow. However, a finite, measurable departure from dimensional scaling persists even at $\gamma=1$. This confirms that VT sustains a residual, anomalous intermittent dynamics. This conclusion is further visually supported by the inset of Fig.~\ref{fig:intermittency}b, which reports the flatness-like ratio $S_6(r)/[S_2(r)]^3$; deviations from the Gaussian value of $15$ strictly persist for $r < \mathcal{L}$ across all evaluated values of $\gamma$. Fig.~\ref{fig:intermittency}d confirms that this anomalous scaling is robust across different Reynolds numbers. Although not explicitly shown, identical qualitative behavior is recovered when employing the NH forcing scheme.

Finally, definitive insight is gained by examining two alternative configurations (Fig.~\ref{fig:intermittency}b): suppressing \textit{only} VT ($\gamma=0.05$) and suppressing the \textit{full} vortex stretching term (VA+VT, $\gamma=1$). Removing VT geometrically aligns the vortex lines, massively enhancing the relative contribution of VA; accordingly, this configuration exhibits increased deviations from dimensional scaling. By stark contrast, the complete suppression of all vortex stretching (VA+VT) leads to a perfect collapse onto the dimensional prediction $\zeta_p/(p\zeta_3) = 1/3$, signaling the absolute disappearance of intermittency. 

Taken together, these results firmly establish the distinct dynamical roles of the two vortex-stretching components. While VA acts as the engine that amplifies turbulent fluctuations, VT functions as a continuous ``tilting engine'' that sustains spatial complexity through geometric reorientation. Consequently, intermittent dynamics are completely eradicated only when both mechanisms are suppressed. Notably, the persistence of anomalous scaling for the $\gamma=1$ limit is also deeply consistent with studies of turbulence driven by steep power-law forcing spectra (recall Fig.~\ref{fig:spectra}c), which have shown that intermittent fluctuations can survive provided the forcing exponent is sufficiently large \citep{biferale-lanotte-toschi-2004}.

\section{Conclusion}
\label{sec:conclusions}

In this work, we have systematically disentangled the distinct dynamical roles of vorticity amplification (VA) and vortex tilting (VT) in three-dimensional homogeneous isotropic turbulence. By selectively suppressing targeted components of the vortex stretching mechanism within the Navier--Stokes equations, we isolated their individual contributions to the energy cascade, the dissipative anomaly, and small-scale intermittency. 

We demonstrated that VA acts as the primary engine for gradient amplification and enstrophy production. As confirmed by our scale-by-scale budget analyses, progressively suppressing VA dismantles the classical forward transfer of kinetic energy, smooths the small-scale structure of the flow, and heavily depletes extreme fluctuation events. In the limit of complete VA suppression ($\gamma=1$), the classical energy dissipative anomaly entirely vanishes, with the normalized energy dissipation decaying strictly as $Re_\lambda^{-1}$, and the system transitions to a purely VT-driven state where enstrophy emerges as the relevant inviscid invariant. Under these conditions, the dynamics are governed by a purely forward enstrophy cascade. Using phenomenological arguments, which we quantitatively validated through spectral fluxes and scaling laws, we predicted and observed a finite asymptotic enstrophy dissipation approached via a $Re_\lambda^{-1/2}$ finite-Reynolds-number correction. Crucially, by driving the flow with both helical (ABC) and non-helical (NH) forcing schemes, we definitively established that these anomalous dissipation limits, alongside the $k^{-3}$ intermediate scaling, are universal, intrinsic properties of the VA-free dynamics, completely independent of the large-scale energy injection mechanism.

Perhaps most remarkably, we found that the flow retains a coherent multiscale organization, robust anomalous scaling, and a broad multifractal spectrum even when VA is completely removed. While suppressing VA significantly narrows the probability density functions of the dissipation, a residual log-normal core and clear deviations from dimensional scaling in the higher-order velocity structure functions persist. This physically isolates the distinct mechanistic roles of the underlying vortex dynamics: while VA amplifies turbulent fluctuations and generates the most extreme gradients, the geometric reorientation provided by VT alone is mechanically sufficient to sustain multifractal intermittent dynamics. 

Taken together, these results highlight a fundamental, mechanistic separation between the processes responsible for gradient amplification and those responsible for geometric complexity and topology in turbulence. This separation provides a rigorous theoretical framework for disentangling the distinct physical origins of anomalous dissipation and intermittency. Ultimately, these insights may inform the development of more accurate reduced-order descriptions of turbulent flows, particularly in the formulation of subgrid-scale models that aim to faithfully replicate the structural complexities of the turbulence cascade.

\section*{Acknowledgments}
The author gratefully acknowledges M. Quadrio and R. K. Singh for fruitful discussions. The author also thanks M. Quadrio for reviewing an earlier version of the manuscript. The author acknowledges CINECA for the availability of high-performance computing resources and support through the ISCRA project KolPar-HP10B89A1O.

\appendix

\section{Inviscid Invariants}
\label{sec:inv}

In this appendix, we derive the inviscid invariants of the modified Navier--Stokes system for the case where VA is suppressed.

We first demonstrate that enstrophy is an exact inviscid invariant of the modified system for $\gamma=1$. In the inviscid limit (i.e., neglecting the viscous term), the modified vorticity equation reads:
\begin{equation}
  \partial_t \omega_i + u_j \partial_j \omega_i = \omega_j \partial_j u_i - \gamma \omega_j \partial_j u_\ell \omega_\ell \omega_i \omega^{-2}.
\end{equation}
Multiplying this equation by $\omega_i$ and exploiting the incompressibility condition ($\partial_j u_j = 0$), we obtain the evolution equation for the enstrophy density:
\begin{equation}
  \frac{1}{2} \partial_t (\omega_i \omega_i) + \frac{1}{2} \partial_j( u_j \omega_i \omega_i ) = ( 1 - \gamma) \omega_i \omega_j \partial_j u_i.
\end{equation}
In the specific case where $\gamma=1$, the right-hand side identically vanishes. Applying the spatial volume average $\langle \cdot \rangle$ and exploiting the properties of a tri-periodic domain (where the integral of a divergence is zero), we obtain:
\begin{equation}
  \frac{\text{d}}{\text{d} t} \langle \omega^2 \rangle = 0.
\end{equation}
This demonstrates that mean enstrophy is an exact inviscid invariant of the system. Consequently, the conservation of the local vorticity magnitude along fluid trajectories implies that all higher-order moments of the enstrophy are also inviscid invariants, i.e.:
\begin{equation}
  \frac{\text{d}}{\text{d} t} \int_{\Omega} |\bm{\omega}|^p \text{d}\Omega = 0 \quad \forall p > 0.
\end{equation}

Conversely, we now demonstrate that helicity is not a general inviscid invariant of the system, even when $\gamma=1$. Defining the mean helicity as $H = \frac{1}{2} \langle u_i \omega_i \rangle$, its time evolution is given by:
\begin{equation}
  \frac{\text{d}H}{\text{d}t} = \frac{1}{2} \langle u_i \partial_t \omega_i + \omega_i \partial_t u_i \rangle.
\end{equation}
Using integration by parts in a periodic domain (which yields $\langle f_i \epsilon_{ijk} \partial_j g_k \rangle = \langle g_i \epsilon_{ijk} \partial_j f_k \rangle$), it is straightforward to show that $\langle u_i \partial_t \omega_i \rangle = \langle \omega_i \partial_t u_i \rangle$. Thus, the evolution equation for $H$ reduces to:
\begin{equation}
  \frac{\text{d}H}{\text{d}t} = \langle u_i \partial_t \omega_i \rangle.
\end{equation}
To substitute $\partial_t \omega_i$ using our modified equation, we recall the vector identity for the curl of a cross product:
\begin{equation}
  \epsilon_{ijk} \partial_j ( \epsilon_{k \ell m} u_\ell \omega_m ) = \omega_j \partial_j u_i - u_j \partial_j \omega_i.
\end{equation}
Using this identity, the helicity evolution becomes:
\begin{equation}
  \frac{\text{d}H}{\text{d}t} = \langle u_i ( - u_j \partial_j \omega_i + \omega_j \partial_j u_i - \gamma \omega_j \partial_j u_\ell \omega_\ell \omega_i \omega^{-2} ) \rangle
   = \langle u_i \epsilon_{ijk} \partial_j ( \epsilon_{k \ell m} u_\ell \omega_m) \rangle - \gamma \langle u_i \omega_j \partial_j u_\ell \omega_\ell \omega_i \omega^{-2} \rangle.
\end{equation}
The first term on the right-hand side vanishes identically. To prove this, we define $V_k = \epsilon_{k \ell m} u_\ell \omega_m$. Applying the product rule, we have:
\begin{equation}
  \langle u_i \epsilon_{ijk} \partial_j V_k \rangle = \partial_j \langle \epsilon_{ijk} u_i V_k \rangle - \langle \epsilon_{ijk} ( \partial_j u_i ) V_k \rangle. 
\end{equation}
The first term vanishes due to spatial homogeneity. For the second term, we exploit the anti-symmetry of the Levi-Civita tensor, noting that $\epsilon_{ijk}(\partial_j u_i) = - \epsilon_{kji}(\partial_j u_i) = - \omega_k$. Therefore:
\begin{equation}
  - \langle \epsilon_{ijk} ( \partial_j u_i ) V_k \rangle = \langle \omega_k V_k \rangle = \langle \omega_k \epsilon_{k \ell m} u_\ell \omega_m \rangle = \langle u_\ell (\epsilon_{k \ell m} \omega_k \omega_m) \rangle = 0.
\end{equation}
This evaluates to zero because the contraction of the anti-symmetric tensor $\epsilon_{k \ell m}$ with the symmetric product $\omega_k \omega_m$ is identically null.

This leaves us solely with the second term in the helicity evolution equation. Let us define $T = u_i \omega_j (\partial_j u_\ell) \omega_\ell \omega_i \omega^{-2}$. We can demonstrate that $\langle T \rangle = 0$ only in the specific case where the flow lacks a net chirality (i.e., when $H = 0$). 
To show this, we introduce a reflected frame of reference such that $x_k' = - x_k$ for all $k$. Consequently, $\partial_k' = - \partial_k$ and $u_k' = - u_k$. The vorticity, being a pseudovector, remains unchanged under reflection: $\omega_k' = \epsilon_{kpq} \partial_p' u_q' = \epsilon_{kpq} ( - \partial_p ) ( - u_q ) = \omega_k$. Evaluating the quantity $T$ in this reflected frame yields:
\begin{equation}
T' = \omega_j' ( \partial_j' u_\ell' ) \omega_\ell' \omega_i' u_i' (\omega')^{-2}
   = \omega_j ( -\partial_j )( - u_\ell ) (\omega_\ell)(\omega_i) (- u_i ) \omega^{-2}
   = - \omega_j ( \partial_j u_\ell) \omega_\ell \omega_i u_i  \omega^{-2} = - T.
\end{equation}
If we assume initial conditions such that $H = 0$, the turbulent field possesses statistical reflectional symmetry. This implies that the probability of finding a certain velocity configuration is identical to that of finding its reflected counterpart, meaning that expected values must be reference-independent ($\langle T' \rangle = \langle T \rangle$). Since $T' = -T$, this requires $\langle T \rangle = - \langle T \rangle$, which forces $\langle T \rangle = 0$. In this scenario, $H$ remains strictly zero.

However, in the case of helical turbulence ($H \neq 0$), reflectional symmetry is broken. Statistically, the velocity field is no longer identical to its reflection, meaning $\langle T \rangle \neq \langle T' \rangle$. Because the symmetry constraint no longer forces the expectation value to zero, we generally have $\langle T \rangle \neq 0$. Consequently, $\text{d}H/\text{d}t \neq 0$. Therefore, helicity is not a general inviscid invariant of this modified system.

\section{Grid Independence and Resolution Requirements}
\label{sec:appendix}

\begin{figure}
  \centering
  \begin{tikzpicture}
  \node at (0,-4.5) {\includegraphics[width=0.7\textwidth]{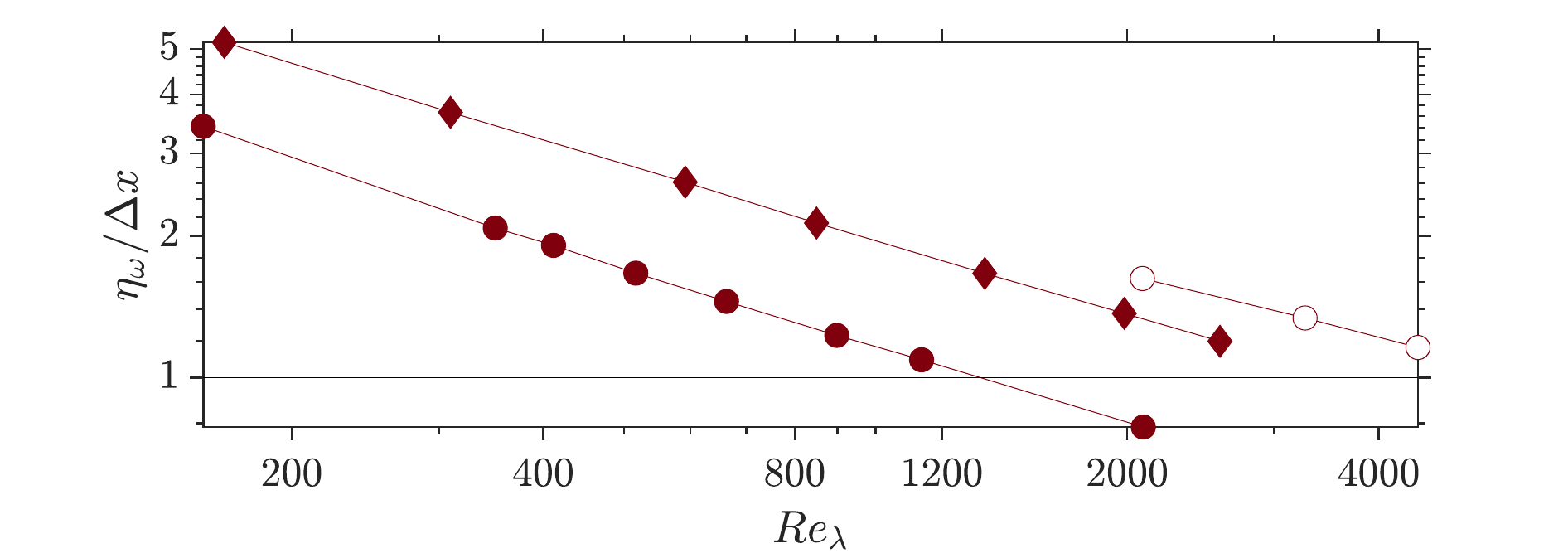}};
  \node at (0, 0.0) {\includegraphics[width=0.7\textwidth]{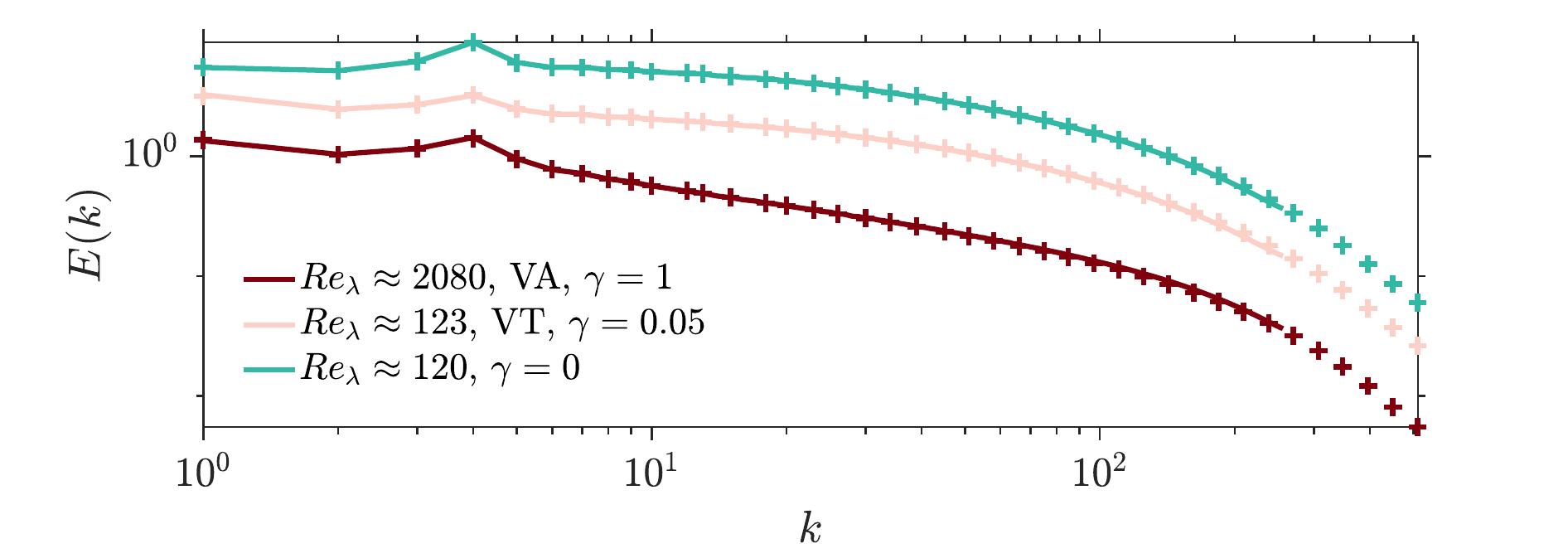}};
  \node at (-5.5,2) {(a)};
  \node at (-5.5,-2.5) {(b)};
  \end{tikzpicture}
\caption{Quantitative assessment of spatial resolution and verification of grid independence. (a) Grid-convergence study comparing the kinetic energy spectra for selected representative configurations. Solid lines represent the $512^3$ baseline grids, while discrete symbols denote the $1024^3$ refined grids. The spectra overlap almost perfectly across all resolved wavenumbers, definitively confirming grid independence. Spectra are vertically shifted for visual clarity. Note that all modified ($\gamma \neq 0$) configurations shown in this panel are driven by the ABC forcing. (b) Evolution of the resolution parameter, $\eta_\omega/\Delta x$, as a function of the Taylor-scale Reynolds number $Re_\lambda$ for the fully VA-suppressed limit ($\gamma=1$). Circles denote simulations driven by the ABC forcing, while diamonds represent the non-helical (NH) forcing. Filled symbols correspond to simulations performed on the baseline $512^3$ grids, while open symbols correspond to the refined $1024^3$ grids. Notice that the spatial resolution strictly satisfies the criterion $\eta_\omega/\Delta x > 1$ across the entire parameter space. To explicitly assess grid convergence, the ABC-forced case at $Re_\lambda\approx2080$ is simulated at both resolutions.}
  \label{fig:conv}
\end{figure}

In this Appendix, we provide a detailed quantitative assessment of the spatial resolution across the various simulation regimes and verify grid independence. 

For the reference Navier--Stokes simulations ($\gamma=0$), the adequacy of the spatial resolution is verified against the classical Kolmogorov length scale, $\eta=\nu^{3/4}\langle \varepsilon \rangle^{-1/4}$. We strictly enforce the condition $\eta/\Delta x=\mathcal{O}(1)$ across all standard cases. Specifically, for simulations on the baseline $512^3$ grid, we obtain $\eta/\Delta x\approx 1.2$ at $Re_\lambda\approx 65$, and $\eta/\Delta x\approx0.7$ at $Re_\lambda\approx120$. To ensure convergence, the $Re_\lambda\approx120$ case was simulated at both $512^3$ and $1024^3$ resolutions; in the refined configuration, the integration time step is correspondingly decreased alongside the spatial spacing to maintain the Courant--Friedrichs--Lewy (CFL) number strictly below unity. As shown in Fig.~\ref{fig:conv}a, the resulting energy spectra exhibit only marginal differences, confirming robust spatio-temporal grid independence. As $\gamma$ increases and VA is suppressed, the nonlinear depletion of small-scale velocity gradients progressively relaxes these traditional resolution requirements.

For the $\gamma=1$ limit, wherein VA is fully suppressed and enstrophy emerges as the relevant inviscid invariant, the dissipative dynamics are no longer governed by the Kolmogorov scale. Instead, the adequacy of the resolution must be assessed using the viscous enstrophy scale \citep{bos-2021}, defined dimensionally as
\begin{equation}
\eta_\omega=\nu^{1/2}\langle \varepsilon_\omega \rangle^{-1/6},
\end{equation}
where $\langle \varepsilon_\omega \rangle$ is the enstrophy dissipation rate. Fig.~\ref{fig:conv}b tracks the resolution ratio $\eta_\omega/\Delta x$ as a function of the Taylor-scale Reynolds number $Re_\lambda$ for the $\gamma=1$ datasets. For simulations utilizing the ABC forcing on the baseline $512^3$ grid, the criterion $\eta_\omega/\Delta x > 1$ is comfortably maintained up to $Re_\lambda\approx 2080$. For larger Reynolds numbers (up to $Re_\lambda\approx4500$), the resolution is increased to $1024^3$ to ensure $\eta_\omega/\Delta x > 1$ remains satisfied in all cases. To definitively verify grid independence, simulation at $Re_\lambda\approx2080$ was performed using both the $512^3$ and $1024^3$ grids. As demonstrated in Fig.~\ref{fig:conv}a, the energy spectra from both resolutions overlap almost perfectly throughout the entire resolved wavenumber range. Furthermore, simulations utilizing the non-helical (NH) forcing scheme were similarly validated; across the entire NH dataset ($Re_\lambda \in [160, 2600]$ on a $512^3$ grid), the resolution is strictly maintained at $\eta_\omega/\Delta x > 1.2$, as explicitly reported in Fig.~\ref{fig:conv}b.

Grid independence was also explicitly verified for the highly restrictive configuration in which only vortex tilting (VT) is suppressed ($\gamma=0.05$). Because suppressing VT promotes the alignment of vortex lines, which strongly favors the generation of small scales, stringent resolution is required. This configuration was simulated on both $512^3$ and $1024^3$ grids, yielding only minor differences between the two (see the energy spectra in Fig.~\ref{fig:conv}a), thus confirming the convergence of the VT-suppressed results.

\end{document}